\documentclass[twocolumn,showpacs,amssymb,aps,prl]{revtex4}

\usepackage{epsfig}
\usepackage{amsfonts}
\usepackage{amssymb}
\usepackage{amsmath}
\usepackage{psfrag}

\newcommand{\mypsfrag}[2]{\psfrag{#1}{\footnotesize{#2}}}
\newcommand{\npsfrag}[3]{\psfrag{#1}[#2]{\footnotesize{#3}}}
\newcommand{\x}{\times}
\newcommand{\dd}{\partial}

\begin{document}

\renewcommand{\thesection}{\arabic{section}}
\newcommand{\alpheqn}{\addtocounter{equation}{1}\setcounter{saveeqn}%
{\value{equation}}\setcounter{equation}{0}%
\renewcommand{\theequation}{\mbox{\arabic{saveeqn}\alph{equation}}}}
\newcommand{\reseteqn}{\setcounter{equation}{\value{saveeqn}}%
\renewcommand{\theequation}{\arabic{equation}}}

\newcommand{\bec}[1]{\mbox{\boldmath $ #1$}}
\renewcommand{\vec}[1]{\mbox{\boldmath $ #1$}}
\newcommand{\Ra}{$Ra\ $}
\renewcommand{\P}{$P\ $}
\newcommand{\Nu}{$Nu\ $}
\renewcommand{\t}{$\tau\ $}
\newcommand{\Pm}{$P_m\ $}
\newcommand{\El}{$\Lambda\ $}

\newcommand{\captionfonts}{\small \sl}

\makeatletter  
\long\def\@makecaption#1#2{%
  \vskip\abovecaptionskip
  \sbox\@tempboxa{{{\bf #1:} \captionfonts #2}}%
  \ifdim \wd\@tempboxa >\hsize
    { {\bf#1:} \captionfonts#2\par}
  \else
    \hbox to\hsize{\hfil\box\@tempboxa\hfil}%
  \fi
  \vskip\belowcaptionskip}
\makeatother   

\newcommand{\figref}[1]{Fig. \ref{#1}}
\newcommand{\Ref}[1]{(\ref{#1})}
\newcommand{\DTc}{\Delta T_{\mathrm{c}}}
\newcommand{\al}{\alpha} 
\newcommand{\eps}{\epsilon} 
\newcommand{\epsBV}{\epsilon_{\mathrm{BV}}}
\newcommand{\epsSBV}{\epsilon_{\mathrm{SBV}}} 
\newcommand{\D}{{\Delta_2}} 
\newcommand{\Hf}{H_{\mathrm{F}}}
\newcommand{\htr}{h_{\mathrm{t}}} 
\newcommand{\nv}{{\bf{\hat  n}}} 
\newcommand{\qv}{{\bf q}} 
\newcommand{\qvc}{{\bf q}_{\mathrm{c}}} 
\newcommand{\qc}{q_{\mathrm{c}}} 
\newcommand{\n}{{\nabla}} 
\newcommand{\kv}{{\bf k}} 
\newcommand{\pt}{{\partial_t}} 
\newcommand{\vv}{{\bf v}} 
\newcommand{\jv}{{\bf j}} 
\newcommand{\iv}{{\bf i}} 
\newcommand{\zv}{{\bf {\hat z}}} 
\newcommand{\zerov}{{\bf 0}} 
\newcommand{\ta}{\Theta} 
\newcommand{\sigT}{\sigma_{\mathrm{T}}}
\newcommand{\ga}{\gamma}

\title{
Planetary Dynamos}
\author{F. H. Busse$^1$ and R. Simitev$^2$\\
{\small \it
$^1$ Physikalisches Institut der Universit\"at Bayreuth, D-95440 Bayreuth, Germany \\
$^2$ Department of Mathematics, University of Glasgow, Glasgow G12 8QW, UK \\
email: busse@uni-bayreuth.de}}

\date{\today}

\begin{abstract}
The theory of planetary dynamos and its applications to observed
phenomena of planetary magnetism are outlined.  
It is generally accepted that convection flows driven by thermal or
compositional buoyancy are the most likely source for the sustenance
of global planetary magnetic fields. 
While the existence of dynamos in electrically conducting fluid
planetary cores provides constraints on properties of the latter, the
lack of knowledge about time dependences of the magnetic fields and
about their toroidal components together with the restricted parameter
regions accessible to theory have prevented so far a full understanding of
the phenomena of planetary magnetism. \\
\textbf{Published in "Planets and Moons", T.~Spohn (ed.), vol.~10 of series
  "Treatise on Geophysics", pp.~281-298, G.~Schubert, (gen.~ed.),
  Elsevier, 2007\\
DOI: 10.1016/B978-044452748-6.00160-7}
\end{abstract} 

\maketitle
\newpage
\section{1. Historical Introduction}
\label{sec:1}

While the Earth's magnetism has been studied for centuries starting
with the first scientific monograph of Gilbert (1600), the question of
the magnetism of other planets had received scant attention until
recently because of the lack of relevant observations. Only in 1955
clear evidence for the existence of a planetary magnetic field other
than the geomagnetic one was obtained through the observation of the
Jovian decametric radio waves (Burke and Franklin, 1955). Since it had
been more or less accepted until the end of the 19-th century that
geomagnetism arises from the remnant  magnetization of the Earth
similar properties may have been assumed for the other terrestrial
planets and the Moon. This view lost its appeal, however, when it
became evident that the Curie-temperature is exceeded in the Earth
below a depth of about 30 km. Ferromagnetic materials in the Earth's
crust could thus explain only magnetic fields with a relatively short
wavelength. 
 
The current period of intense research on the magnetism of planets
other than that of the Earth started with the first detailed
measurement of Jupiter's magnetic field by the Pioneer 10 space probe
in 1973 and the discovery of Mercury's magnetism by Mariner 10 in
1974. In the early seventies also the development of the theory of
magnetohydrodynamic dynamos had started in which the reaction of the
Lorentz force of the generated magnetic field is taken into account in
physically realistic configurations (Childress and Soward, 1972;
Busse, 1973; Soward, 1974). Until that time dynamo theoreticians had
focused their attention on the kinematic problem in which the
possibility of growing magnetic fields driven by somewhat arbitrarily
chosen velocity fields is considered. It must be remembered that only
a few years earlier it had been demonstrated by Backus (1958) and
Herzenberg (1958) in a mathematically convincing way that the
homogeneous dynamo process of the generation of magnetic fields in a
singly connected domain of an electrically conducting fluid is indeed
possible. Doubts about the feasibility of this process which had first
been proposed by Larmor (1919)  had persisted after Cowling (1934) had
proved that purely axisymmetric fields could not be generated in this
way. 

The complexity of the magnetohydrodynamic dynamo problem described by
the nonlinearly coupled Navier-Stokes equations and the equation of
magnetic induction had prevented progress in understanding planetary
dynamos through analytical solutions. Only the advent of powerful
enough computers in the 1990-ies has allowed to solve numerically the
coupled three-dimensional partial differential equations through
forward integration in time. Even today and for the foreseeable future
the limits of computer capacity will permit the exploration of only a
small fraction of the parameter space of interest for the
understanding of planetary dynamos. 

In view of the difficulties of a rigorous theory of planetary dynamos,
many attempts have been made to obtain simple similarity relationships
which would fit the observed planetary magnetic moments as function of
certain properties of the planets. Some early proponents have gone as
far as claiming the existence of a "magnetic Bode's law" corresponding
to a relationship between the magnetic moment and size or angular
momentum of a planet in analogy to the Titius-Bode law for the radii
of the orbits of the planets. Just as in the latter case, however,
attempts to derive a magnetic Bode's law from basic physical
principles have failed. 

Other proposals have taken into account physical forces. Since a
common ingredient of planetary dynamos is the existence of a fluid
part of the core with a sufficiently high electrical conductivity the
latter parameter together with the core radius and the angular
velocity of the planetary rotation usually enter the similarity
relationships such as those proposed by Hide (1974), Busse (1976),
Jacobs (1979) and Dolginov (1977). Malkus (1968, 1996) has argued for
the  precession of the Earth as the cause of geomagnetism and he and
Vanyo (1984) have demonstrated through laboratory experiments that
precession and tides may cause turbulent motions in fluid planetary
cores. Dolginov (1977) proposed a scaling law for the precessional
origin of all planetary magnetic fields. While a dynamo driven by
turbulent flows caused by precession and tides can not be easily excluded
in the case of the Earth (Tilgner, 2005), it is much less likely in
the case of other planets such as Uranus for which precessional
torques are rather minute. Just as a common precessional origin of
planetary magnetism is not  regarded as feasible, so have all other
proposed similarity relationships lost in appeal  and are  no longer
seriously considered. We shall return, however, to scaling
relationships based more directly on the basic equations in section
6. 

\section{2. General Remarks on the Dynamo Theory of Planetary Magnetism}
\label{sec:2}

Since the proposal of the geodynamo as the cause of the Earth's
magnetism had been in doubt for a long time before 1958, numerous
alternative proposals had been made in the literature. Among these 
only the possibility that thermoelectric currents may generate a
planetary magnetic field is still discussed in the case of Mercury
(Stevenson, 1987; Giampieri and Balogh, 2002). For a discussion of the
failings of the various proposals for non-dynamo origins of planetary
magnetic fields we refer to the papers of Bullard (1949) and Stevenson
(1983).  
Although the dynamo hypothesis of the origin of planetary magnetism is
not without difficulties, it is the only one considered seriously at
the present time with the possible exception of the just mentioned
case of Mercury. 

Dynamos generally convert mechanical energy into magnetic one. In
contrast to the technical dynamo which is characterized by a multiply
connected region of high electrical conductivity, i.e. it depends on
an appropriate wiring, planetary dynamos are referred to as
homogeneous dynamos since they operate in a singly connected domain of
high electrical conductivity. Since flows in planetary cores with
active dynamos are usually turbulent the small scale structure of the
magnetic field is correspondingly chaotic. 
The large scale structure, however, can be quite regular. One
distinguishes "steady" and oscillatory dynamos. The most famous
example of the latter kind is the solar dynamo which exhibits a well
defined period of about 22 years. The geodynamo, on the other hand, is
a "steady" dynamo, even though it varies in its amplitude by a factor of
two or more on the magnetic diffusion time scale and reverses its
polarity on a longer time scale. A measure of the magnetic diffusion
time is given by the decay time, $t_d = \sigma \mu r_0^2/\pi^2$, of
the magnetic field in the absence of fluid motions. Here $\sigma$ and
$\mu$ refer to the electrical conductivity and the magnetic
permeability of the planetary core of radius $r_0$. In the case of the
Earth the decay time is of the order of 20 000 years, but it may vary
between a few hundred and a million years for other examples of
planetary dynamos. 

The theory of homogeneous dynamos is based on Maxwell's equations for
the magnetic flux density $\vec B$, the electric current density $\vec
j$ and the electric field $\vec E$ in the magnetohydrodynamic  
approximation in which the displacement current is neglected. This
approximation is highly accurate as long as the fluid velocity is
small compared to the  
velocity of light which is certainly the case for all planetary
applications. These equations together with Ohm's law for a moving  
conductor are given by
\begin{subequations}
\begin{align}
\nabla\cdot \vec B &= 0,\qquad    \frac{\partial}{\partial t}\vec B =
- \nabla\times\vec E,   \qquad \\ \nabla\times\vec B &= \mu\vec j,
\qquad  \vec j = \sigma (\vec v\times\vec B + \vec E), 
\end{align}
\end{subequations}
where $\mu$ is the magnetic permeability of the fluid and $\sigma$ is
its electrical conductivity. 
These "pre-Maxwell" equations have the property that they are
invariant with respect to a Galilei 
transformation, i. e. the equations remain unchanged in a new frame of reference
moving with the constant velocity vector $\vec V$ relative to the
original frame of reference. Indicating the variables of the new frame
by a prime we find 
\begin{subequations}
\begin{align}
\vec v' &= \vec v - \vec V, \qquad  \frac{\partial}{\partial t'} = \frac{\partial}{\partial t}  +  \vec V \cdot \nabla ,\qquad \\
\vec B' &= \vec B,\qquad     \vec E' = \vec E  +  \vec V \times \vec B,\qquad    \vec j' = \vec j\; .
\end{align}
\end{subequations}
This invariance is the basis for the combination in MHD of equations
(1) with the equations of hydrodynamics in their usual
non-relativistic form. It is remarkable that this invariance does not only
hold with respect to a Galilei transformation, but with respect to a
transformation to a rotating frame of reference as well. In that case
$\vec V$ is replaced by $\vec \Omega \times \vec r$ in equations (2),
but when $\frac{\partial}{\partial t'}$ is operating on any vector
$\vec a$ the term $-\vec \Omega \times \vec a$ must be added on the
right hand side, since even a constant vector field becomes time
dependent when seen from a rotating frame unless it is parallel to
$\vec \Omega$.

Elimination of $\vec E$ and $\vec j$ from equation (1) yields the
equation of magnetic induction 
\begin{equation}
\nabla \times \left(\frac{1}{\sigma\mu} \nabla \times\vec B\right) \\= \frac{\partial}{\partial t}\vec B + \nabla \times ( \vec v \times\vec B )
\end{equation}
which for a solenoidal velocity field $\vec v$ and a constant magnetic diffusivity $\lambda \equiv (\sigma\mu)^{-1}$ can be further simplified,
\begin{equation}
\left(\frac{\partial}{\partial t} +\vec v \cdot \nabla\right)\vec B - \lambda \nabla^2\vec B \\= \vec B \cdot\nabla \vec v .
\end{equation}
This equation has the form of a heat equation with the magnetic field
line stretching term on the  
right hand side acting as a heat source. This interpretation is
especially useful for the dynamo problem. In order that  a magnetic
field $\vec B$ may grow, the term on the right hand side of (4) must
overcome the effect of the  
magnetic diffusion term on the left hand side. Using a typical
velocity $U$ and a typical length 
scale $d$, the ratio of the two terms can be estimated by the magnetic
Reynolds number $Rm$, 
\begin{equation}
U d/\lambda \equiv Rm .
\end{equation}
Only when $Rm$ is of the order one or larger may growing magnetic fields become possible.

In the following we shall first consider the mathematical formulation
of the problem of convection driven dynamos in rotating spherical
shells in a simple form in which only the physically most relevant
parameters are taken into account. Before discussing dynamo solutions
in section 5 we shall briefly outline in section 4 properties of
convection it the absence of a magnetic field. Applications to various
planets and moons will be considered in section 6 of this article and
some concluding remarks are given in section 7.  
%
%
\begin{figure}
\mypsfrag{z}{$\vec k$}
\mypsfrag{x}{\hspace*{-10mm}$\varphi=0$}
\mypsfrag{r}{$\vec r$}
\mypsfrag{th}{$\theta$}
\mypsfrag{phi}{$\varphi$}
\mypsfrag{om}{$\vec \Omega$}
\mypsfrag{d}{$d$}
\begin{center}
\hspace*{0mm}
\epsfig{file=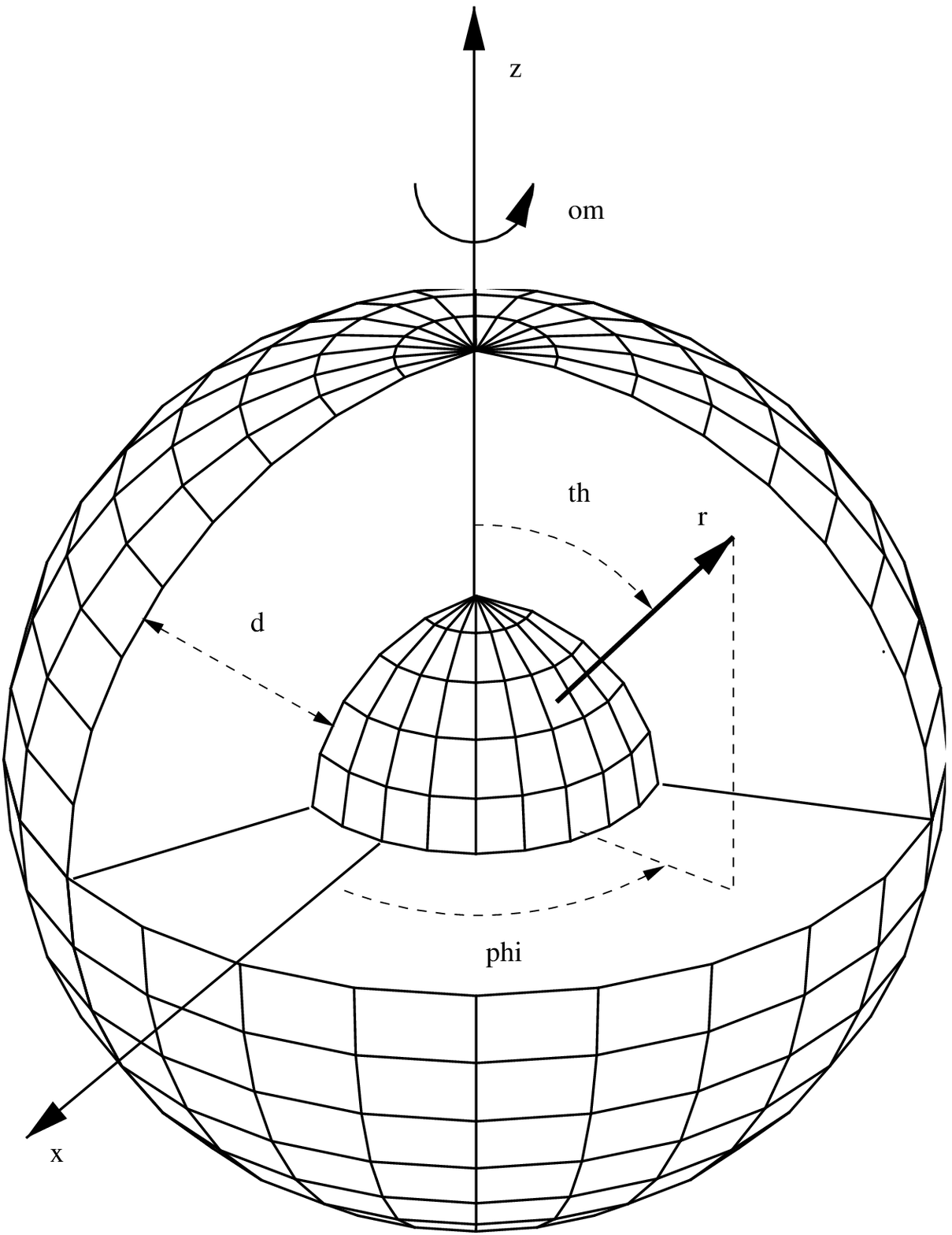,height=6cm,clip=}
\end{center}
\caption[]{Geometrical configuration of the problem. A part of the
 outer spherical surface is removed to expose the interior of the
 shell to which the conducting fluid is confined.}
\label{f.01}
\end{figure}

\section{3. Mathematical formulation of the problem of spherical dynamos}
\label{sec:3}

A sketch of the geometrical configuration that will be considered is
shown in figure 1. For the equations describing convection driven
dynamos in the frame rotating with the angular velocity $\Omega$ 
we use a standard formulation which has also been used for a dynamo
benchmark (Christensen \emph{et  al.} 2001). But we assume different
scales and assume that a more general  
static state exists with the temperature distribution $T_S = T_0 - \beta
d^2 r^2 /2 + \Delta T \eta r^{-1} (1-\eta)^{-2}$ where $\eta$ denotes
the ratio of inner to outer radius of the spherical shell and $d$ is its thickness.
$\Delta T$ is the temperature difference between the boundaries in the
special case $\beta =0$. In the case $\Delta T = 0$ the static temperature distribution $T_S$ corresponds to that of a homogeneously heated sphere with the heat source density proportional to the parameter $\beta$.
The gravity field is given by $\vec g = - \gamma d \vec r$ where
$\vec r$ is the position vector with respect to the center of the sphere and
$r$ is its length measured in units of $d$. 

In addition to  $d$, the
time $d^2 / \nu$,  the temperature $\nu^2 / \gamma \alpha d^4$ and 
the magnetic flux density $\nu ( \mu \varrho )^{1/2} /d$ are used as
scales for the dimensionless description of the problem  where $\nu$ denotes
the kinematic viscosity of the fluid, $\kappa$ its thermal diffusivity and
$\varrho$ its density.
The Boussinesq approximation is used in that $\varrho$ is assumed to be
constant except in the gravity term where its temperature dependence given by
$\alpha \equiv - ( d \varrho/dT)/\varrho =$ const. is taken into
account. The dimensionless equations of motion, the heat equation for
the deviation $\Theta$ of the temperature field from the static
distribution and the equation of magnetic induction thus assume the
form 
\begin{subequations}
\begin{align} 
&\n^2\vec v + \vec B \cdot \nabla \vec B + \vec r \ta - \n \pi \nonumber \\&\hspace*{1cm} = P^{-1}(\partial_t\vec v + \vec v\cdot\n \vec v) 
+ \tau \kv \times \vec v \\ & \n \cdot \vec v = 0\\ &\nabla^2 \Theta + \left[ R_i +R_e \eta r^{-3} (1 - \eta)^{-2} \right]\vec r \cdot \vec v \nonumber \\ &\hspace*{1cm}= P ( \partial_t + \vec v \cdot \nabla ) \Theta\\
&P_m\left(\frac{\partial}{\partial t} +\vec v \cdot \nabla\right)\vec B - \nabla^2\vec B  = P_m\vec B \cdot\nabla \vec v
\end{align}
\end{subequations} 
where $\kv$ is the unit vector in the direction of the axis of rotation and where $\n \pi$ includes all terms that can be written as gradients. The Rayleigh numbers $R_i$ and $R_e$, the Coriolis parameter $\tau$, the Prandtl
number $P$ and the magnetic Prandtl number $P_m$ are defined by
\begin{subequations}
\begin{align}
& R_i = \frac{\alpha \gamma \beta d^6}{\nu \kappa} , 
\enspace R_e = \frac{\alpha \gamma \Delta T d^4}{\nu \kappa} ,
\\ & \tau = \frac{2
\Omega d^2}{\nu} ,\enspace P = \frac{\nu}{\kappa} , \enspace P_m = \frac{\nu}{\lambda}
\end{align}
\end{subequations}
For simplicity $R_e=0$ will be assumed unless indicated otherwise. The
notation $R \equiv R_i$ will thus be used in the following. The
Prandtl number $P$ has been added as an important parameter of the
problem even though $P=1$ is often assumed with the argument that all
effective diffusivities are equal in turbulent media. The effective
diffusivities of scalar and vector quantities in turbulent fluid flow
differ in general, however, and large differences in the corresponding
molecular diffusivities will not be erased entirely in the turbulent
case. 

Since the velocity field $\vec v$ as well as the magnetic flux density 
$\vec B$ are solenoidal vector fields,  
the general representation in terms of poloidal and toroidal
components can be used,
\begin{subequations}
\begin{align} 
\vec v &= \nabla \times ( \nabla \Phi \times \vec r) + \nabla \Psi \times 
\vec r \enspace ,\\
\vec B &= \nabla \times  ( \nabla h \times \vec r) + \nabla g \times 
\vec r \enspace .
\end{align}
\end{subequations} 
By multiplying the (curl)$^2$ and the curl of equation (6a)
by $\vec r$ we obtain two equations for $\Phi$ and $\Psi$
\begin{subequations}
\begin{align} 
&[( \nabla^2 - \partial_t) L_2 + \tau \partial_{\phi} ] \nabla^2 \Phi + \tau Q
\Psi - L_2 \Theta \nonumber \\ 
&\hspace*{1cm}= - \vec r \cdot \nabla \times [ \nabla \times ( \vec v \cdot
\nabla \vec v - \vec B \cdot \nabla \vec B)]\\
&[( \nabla^2 - \partial_t) L_2 + \tau \partial_{\phi} ] \Psi - \tau Q\Phi \nonumber \\
&\hspace*{1cm}= \vec
r \cdot \nabla \times ( \vec v \cdot \nabla \vec v - \vec B \cdot \nabla \vec B)
\end{align}
\end{subequations} 
where $\partial_t$ and $\partial_{\phi}$ denote the partial derivatives
with respect to time $t$ and with respect to the  
angle $\phi$ of a spherical system of coordinates $r, \theta, \phi$
and where the operators $L_2$ and $Q$ are defined by 
\begin{displaymath}
L_2 \equiv - r^2 \nabla^2 + \partial_r ( r^2 \partial_r)
\end{displaymath}
\begin{displaymath}
Q \equiv r \cos \theta \nabla^2 - (L_2 + r \partial_r ) ( \cos \theta
\partial_r - r^{-1} \sin \theta \partial_{\theta})
\end{displaymath}
The equations for $h$ and $g$ are obtained through the multiplication of
equation (6d) and of its curl by $\vec r$,
\begin{subequations}
\begin{align} 
\nabla^2 L_2 h &= P_m [ \partial_t L_2 h - \vec r \cdot \nabla \times ( \vec v
\times \vec B )]\\
\nabla^2 L_2 g &= P_m [ \partial_t L_2 g - \vec r \cdot \nabla \times ( \nabla
\times ( \vec v \times \vec B ))]
\end{align}
\end{subequations} 
Either rigid boundaries with fixed temperatures as in the benchmark case (Christensen \emph {et al.} 2001),
\begin{displaymath}
\Phi = \partial_{r}(r\Phi) = \Psi = \Theta = 0 \enspace \mbox{ at }
\enspace r=r_i \equiv \eta / (1- \eta) 
\end{displaymath}
\begin{equation}
\enspace \mbox{ and at } \enspace r=r_o =
(1-\eta)^{-1},
\end{equation}
or stress-free boundaries
with fixed temperatures,
\begin{displaymath}
\Phi = \partial^2_{rr}\Phi = \partial_r (\Psi/r) = \Theta = 0 \enspace \mbox{ at }
\enspace r=r_i \equiv \eta / (1- \eta) 
\end{displaymath}
\begin{equation}
\enspace \mbox{ and at } \enspace r=r_o =
(1-\eta)^{-1}
\end{equation}
are often used. The latter boundary conditions are assumed in the
following since they allow to cover numerically a larger region of the
parameter space. The case $\eta = 0.4$ will be considered 
unless indicated otherwise. It provides a good compromise for the
study of both, the regions inside and outside the tangent
cylinder. The latter is the cylindrical surface touching the inner
spherical boundary at its equator. 
For the magnetic field electrically insulating
boundaries are used such that the poloidal function $h$ must be 
matched to the function $h^{(e)}$ which describes the  
potential fields 
outside the fluid shell  
\begin{equation}
g = h-h^{(e)} = \partial_r ( h-h^{(e)})=0 \; 
\mbox{ at } r=r_i \mbox{ and } r=r_o .
\end{equation}
But computations for the case of an inner boundary with no-slip
conditions and an electrical conductivity equal to that of the fluid
have also been done.
The numerical integration of equations (2) together with boundary
conditions (4) proceeds with the pseudo-spectral 
method as described by Tilgner and Busse (1997) and Tilgner (1999)
which is based on an expansion of all dependent variables in
spherical harmonics for the $\theta , \phi$-dependences, i.e. 
\begin{equation}
\Phi = \sum \limits_{l,m} V_l^m (r,t) P_l^m ( \cos \theta ) \exp \{ im \phi \}
\end{equation}
and analogous expressions for the other variables, $\Psi, \Theta, h$ and $g$. 
$P_l^m$ denotes the associated Legendre functions.
For the $r$-dependence expansions in Chebychev polynomials are used. 

For the computations to be reported in the following a minimum of
33 collocation points in
the radial direction and spherical harmonics up to the order 64 have been
used. But in many cases the resolution was increased to 49 collocation
points and spherical harmonics up to the order 96 or 128.

It should be emphasized that the static state $\vec v = \vec B = \Theta = 0$ represents a solution of equations (6) for all values of the Rayleigh numbers $R_i$ and $R_e$, but this solution is unstable except for low or negative values of the latter parameters. Similarly, there exist solutions with $\vec B = 0$, but $\vec v \not= 0, \Theta \not= 0$, for sufficiently large values of either $R_i$ or $R_e$ or both, but, again, these solutions are unstable for sufficiently large values of $P_m$ with respect to disturbances with $\vec B \not= 0$. Dynamo solutions as all solutions with $\vec B \not= 0$ are called are thus removed by at least two bifurcations from the basic static solution of the problem. 

Finally we present in table 1 a list of the most important
parameters used in the following sections.
\begin{table}
\begin{tabular}{l@{\extracolsep{5mm}}l@{\extracolsep{5mm}}l}\hline
$E$ &     Average density of kinetic energy  &   (18)\\
$E_p$, $E_t$ & Energy densities of poloidal & \\
 & and toroidal components of motion & (18)\\
$\eta$ &   Radius ratio of fluid shell & \\
$\Lambda$ & Elsasser number     &  (20)\\
$M$ &       Average magnetic energy density  &    (19) \\
$M_p$, $M_t$  & Energy densities of poloidal and toroidal & \\
 & components of the magnetic field&  (19) \\
$Nu$   &   Nusselt number                          &      (16)\\
$P$    &    Prandtl number                    &              (7b)\\
$P_m$  & Magnetic Prandtl number             &   (7b)\\
$R$    &    Rayleigh number                           &    (7a)\\
$R_c$  &  Critical value of $R$ for onset of convection & \\
$Rm$   &  Magnetic Reynolds  number $= P_m\sqrt{2E}$ & \\
$\tau$ &    Coriolis parameter  &                           (7a) \\[2mm]\hline
\end{tabular}\\[2mm]
\caption[]{Important dimensionless parameters.}
\end{table}

\section{4. Convection in rotating spherical shells}
\label{sec:4}

For an introduction to the problem of convection in spherical shells 
we refer to the review of Busse (2002a). Convection tends to set in
first outside the tangent cylinder in the form of thermal Rossby waves
for which the Coriolis force is balanced almost entirely by the
pressure gradient. The model of the 
rotating cylindrical annulus has been especially useful for the
understanding of this type of convection.
A rough idea of the dependence of the critical Rayleigh number $R_{c}$
for the onset of convection on
the parameters of the problem can be gained
from the expressions derived from the annulus model (Busse, 1970)
\begin{subequations}
\begin{align} 
R_{c} = 3 \left( \frac{P \tau }{1+P} \right)^{\frac{4}{3}} ( \tan
\theta_m)^{\frac{8}{3}} r_m^{-\frac{1}{3}} 2^{-\frac{2}{3}}\\   
m_c = \left( \frac{P \tau}{1+P} \right)^{\frac{1}{3}} ( r_m \tan
\theta_m )^{\frac{2}{3}} 2^{-\frac{1}{6}} \\ 
\omega_c = \left( \frac{\tau^2}{(1+P)^2P} \right)^{\frac{1}{3}}
2^{-\frac{5}{6}} 
(\tan^2 \theta_m / r_m )^{\frac{2}{3}}
\end{align}
\end{subequations}
where $r_m$ refers to the mean radius of the fluid shell, $r_m = (r_i + r_o)/2$,
and $\theta_m$ to the corresponding colatitude, $\theta_m =$ arcsin $(r_m(1-\eta))$.
The azimuthal wavenumber of the preferred mode is denoted by $m_c$ and the
corresponding angular velocity of the drift of the convection columns in the
prograde direction is given by $\omega_c / m_c$. 
\begin{figure}[ht]
\mypsfrag{a0}{$10^0$}
\mypsfrag{a1}{$10^1$}
\mypsfrag{a2}{$10^2$}
\mypsfrag{a3}{$10^3$}
\mypsfrag{a4}{$10^4$}
\mypsfrag{a5}{$10^5$}
\mypsfrag{a6}{$10^6$}
\mypsfrag{a7}{$10^7$}
\mypsfrag{b4}{$10^4$}
\mypsfrag{0.1}{$0.1$}
\mypsfrag{1.0}{$1$}
\mypsfrag{10.0}{$10$}
\mypsfrag{100.0}{$100$}
\mypsfrag{R}{$R_c$}
\mypsfrag{w}{$\omega_c$}
\mypsfrag{P}{$P$}
\begin{center}
\hspace*{-4mm}
\epsfig{file=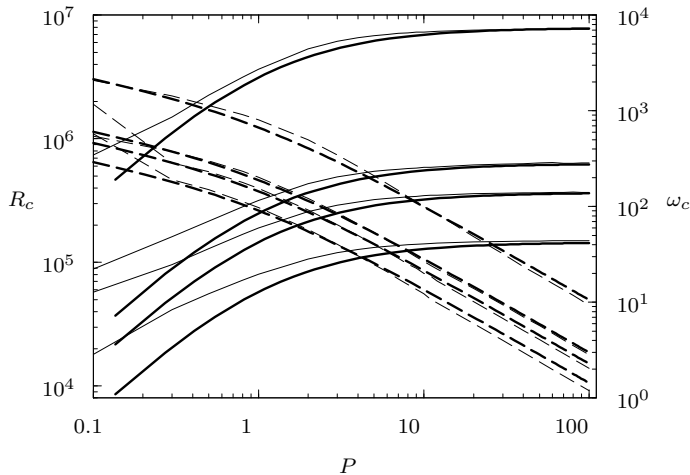,height=9cm,angle=-90}
\end{center}
\caption{Critical Rayleigh number $R_{c}$ (thin solid lines) and
  frequency $\omega_c$ (right ordinate, thin dashed lines) as a
  function of the Prandtl number $P$ in the case $\eta = 0.4$ for the
  Coriolis numbers $\tau = 5 \times 10^3$, $10^4$, $1.5 \times 10^4$
  and $10^5$ (from bottom to top). The thick lines correspond to
  expressions (15a) and   (15c).}
\label{f2}
\end{figure}

In figure 2 expressions (15a,c) are compared with accurate numerical
values which indicate that the general  trend is well represented by
expressions (15a,c). 
The same property holds for $m_c$. For a rigorous asymptotic analysis
in the case $P = 1$ 
including the radial dependence we
refer to Jones {\it et al.} (2000). 
\begin{figure}[ht]
\begin{center}
\epsfig{file=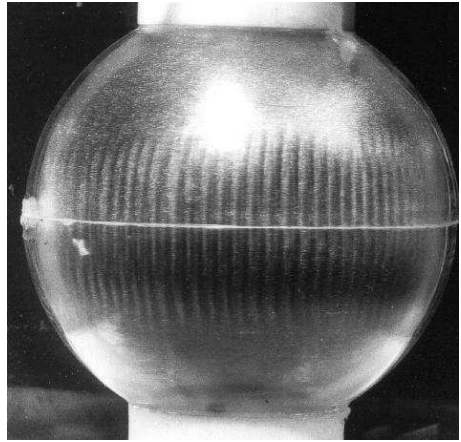,width=6cm}
\end{center}
\caption{Banana cells in a thin rotating spherical fluid shell cooled
from within. Convection driven by centrifugal buoyancy is made visible
by a suspension of tiny flakes which become aligned with the
shear. (After Busse and Carrigan, 1976)} 
\label{f3}
\end{figure}

It is evident from figure 2 that the agreement between expressions
(15) and the numerical values deteriorates as low values of $P$ are
approached. This behavior is caused by the fact that instead of the thermal Rossby wave mode the inertial
mode of convection becomes preferred at onset 
for sufficiently low Prandtl numbers. It is characterized by convection
cells attached to the equatorial region of the outer boundary not unlike
the ``banana cells'' seen in the narrow gap experiment of figure 3
. The equatorially attached convection does indeed represent an inertial wave
modified by the effects of viscous dissipation and thermal
buoyancy. An analytical description of this type of convection can
thus be attained through the introduction of viscous friction and
buoyancy as perturbations as has been done by Zhang(1994) and by Busse and Simitev (2004) for  stress-free and by Zhang (1995) for no-slip boundaries. According to Ardes {\it et al.} (1997) equatorially attached
convection is preferred at onset for $\tau < \tau_l$ where $\tau_l$
increases in proportion to $P^{-1/2}$.

A third form of convection is realized in the polar regions of the
shell which comprise the two fluid domains inside the tangent
cylinder. Since gravity and rotation vectors are nearly parallel in
these regions (unless values of $\eta$ close to unity are used) convection
resembles the kind realized in a horizontal layer heated from below
and rotating about a
vertical axis. Because the Coriolis force can not be balanced by the pressure gradient in this case, the onset of convection is delayed to higher values of $R$ where convection outside the tangent cylinder has reached already high amplitudes. In the case of $\eta = 0.4$ the onset of convection in the polar regions typically occurs at
Rayleigh numbers which exceed the critical values $R_c$ for
onset of convection outside the tangent cylinder by a factor of the order $4$. 
Except for the case of very low Prandtl numbers the retrograde differential rotation in the polar regions generated by convection outside the tangent cylinder tends
to facilitate polar convection by reducing the rotational constraint.
A tendency towards an alignment of convection rolls
with the North-South direction (Busse and Cuong 1977) can be noticed,
but this property is superseded by instabilities of the
K\"uppers-Lortz type (for an experimental demonstration see Busse and Heikes (1980)) and by interactions with turbulent convection
outside the tangent cylinder. 

\begin{figure}[t]
\begin{center}
\hspace*{0cm}
\epsfig{file=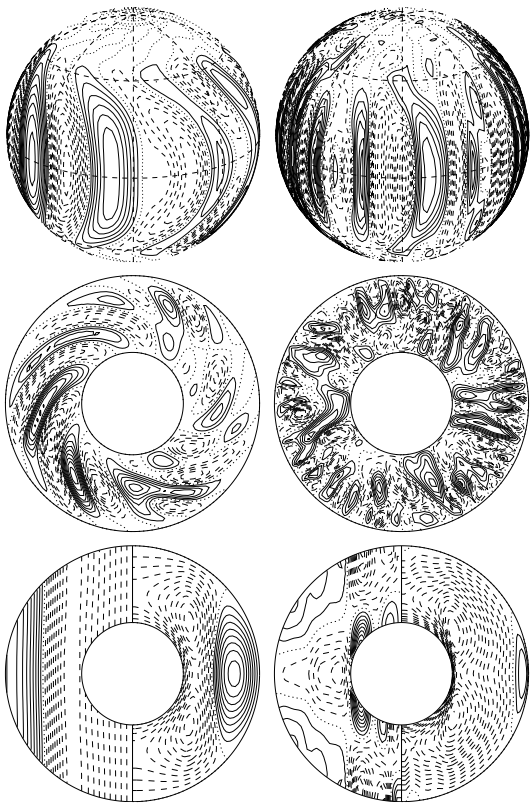,width=7cm,clip=}
\end{center}
\caption{Convection in rotating spherical fluid shells in the cases
  $\tau = 10^4$, $R=4 \times 10^5$, $P=1$ (left column) and $\tau =  5
  \times 10^3$, $R= 8 \times 10^5$, $P=20$  (right column). Lines of
  constant $u_r$ in the middle spherical surface, $r=r_i +0.5$
  are shown in the upper row. The plots of the middle row show
  streamlines, $r \partial \Phi / \partial \varphi=$ const., in the
  equatorial plane. The lowermost plots indicate lines of constant
   mean azimuthal velocity $\bar u_{\varphi}$ in the left halves
  and isotherms of $\bar{\Theta}$   in the right halves.  }
\label{f4}
\end{figure}
Typical features of low and high Prandtl number convection are
illustrated in figure 4. The columnar nature of convection does not
vary much with $P$ as is evident from the top two plots of the
figure. At Prandtl numbers of the order unity or less, - but not in
the case of inertial convection-, the convection columns tend to
spiral away from the axis and thereby create a Reynolds stress which
drives a strong geostrophic differential rotation as shown in the
bottom left plot of the figure. This differential rotation in turn
promotes the tilt and a feedback loop is thus created. At high values
of $P$ the Reynolds stress becomes negligible and no significant tilt
of the convection columns is apparent. In this case the differential
rotation is generated in the form of a thermal wind caused by the
latitudinal gradient of the axisymmetric component of $\Theta$. 
\begin{figure}
\psfrag{0}{0}
\psfrag{30}{30}
\psfrag{60}{60}
\psfrag{90}{90}
\psfrag{120}{120}
\psfrag{150}{150}
\psfrag{180}{180}
\psfrag{0}{0}
\psfrag{1}{1}
\psfrag{10}{10}
\psfrag{Nui}{$Nu_i$}
\psfrag{Nuo}{$Nu_o$}
\psfrag{th}{$\theta$}
\begin{center}
\hspace*{-8mm}
\epsfig{file=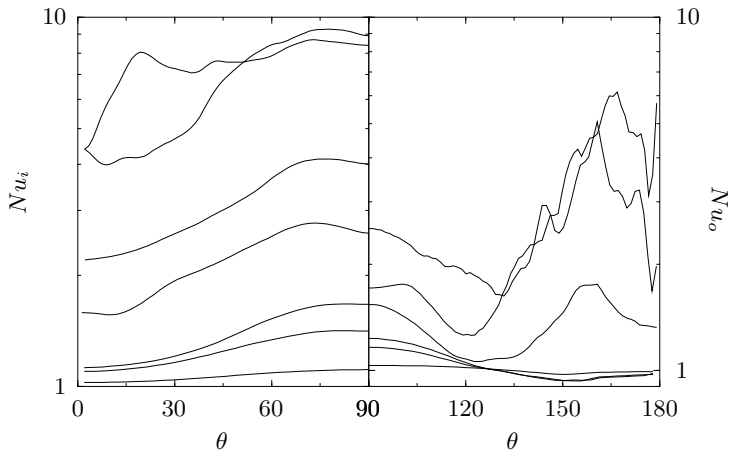,height=9.5cm,angle=-90,clip=}
\end{center}
\caption{The time- and azimuthally-averaged local Nusselt numbers at
  the inner spherical boundary $Nu_i$ (left plot) and
  at the outer spherical boundary $Nu_o$ (right plot) as functions of
  the colatitude $\theta$ for  $P=0.5$,
  $\tau=1.5\times10^4$ and $R= 4,8,10,15,20,30,25\times10^5$
  (from bottom to top at $90^\circ$). }
\label{f5}
\end{figure}

Among the properties of convection at finite amplitudes the heat
transport is the most important one. Customarily its efficiency is
measured by the Nusselt number which is defined as the heat transport
in the presence of convection divided by the heat transport in the
absence of motion. In the case of the spherical fluid shell two
Nusselt numbers can be defined measuring the efficiency of convection
at the inner and the outer boundary, 
\begin{equation}
Nu_i=1- \frac{P}{r_i} \left.\frac{d \overline{\overline{\Theta}}}{d r}\right|_{r=r_i}\qquad Nu_o=1- \frac{P}{r_o} \left.\frac{d \overline{\overline{\Theta}}}{d r}\right|_{r=r_o}
\end{equation}  
where the double bar indicates the average over the spherical surface. In addition local Nusselt numbers
\begin{equation}
Nu_i(\theta) =1- \frac{P}{r_i} \left.\frac{d\overline{\Theta}}{d r}\right|_{r=r_i}\qquad Nu_o(\theta) =1- \frac{P}{r_o} \left.\frac{d\overline{\Theta}}{d r}\right|_{r=r_o}
\end{equation}  
are of interest where only the azimuthal average is applied, as
indicated by the single bar, instead of the average over the entire
spherical surface. Examples of such measures of the dependence of the
heat transport on latitude are shown in figure 5. This figure
demonstrates that at low supercritical Rayleigh numbers the heat
transport occurs primarily across the equatorial region, but as $R$
increases the heat transport in the polar regions takes off and soon
exceeds that at low latitudes. This process is especially evident at
the outer boundary. In the polar regions convection is better adjusted
for carrying heat from the lower boundary to the upper one, and it is
known from computations of the convective heat transport in horizontal
layers rotating about a vertical axis that the value of $Nu$ may
exceed the value in a non-rotating layer at a given value of $R$ in
spite of the higher critical Rayleigh number in the former case
(Somerville and Lipps, 1973). 

Outside the tangent cylinder the convective heat transport encounters
unfavorable conditions in that the cylindrical form of the convection
eddies is not well adjusted to the spherical boundaries. This handicap
is partly overcome through the onset of time dependence in the form of
vacillations in which the convection columns expand and contract in
the radial direction or vary in amplitude. 

At Prandtl numbers of the order unity and less another effect
restricts the heat transport. The shear of the geostrophic
differential rotation created by the Reynolds stresses of the
convection columns severely inhibits the heat transport. To illustrate
this effect we have plotted in figure 6 in addition to the Nusselt
numbers the averages of the kinetic energy densities of the various
components of the convection flow which are defined by
\begin{subequations}
\begin{align}
&
\overline{E}_p = \frac{1}{2} \langle \mid \nabla \times ( \nabla \bar \Phi \times \vec r )
\mid^2 \rangle , \quad \overline{E}_t = \frac{1}{2} \langle \mid \nabla \bar \Psi \times
\vec r \mid^2 \rangle, \\
&
\check{E}_p = \frac{1}{2} \langle \mid \nabla \times ( \nabla \check \Phi \times \vec r )
\mid^2 \rangle , \quad \check{E}_t = \frac{1}{2} \langle \mid \nabla \check \Psi \times
\vec r \mid^2 \rangle,
\end{align}
\end{subequations}
where the angular brackets indicate the average over the fluid shell
and $\bar \Phi$ refers to the azimuthally averaged component of $\Phi$,
while $\check \Phi$ is defined by $\check \Phi = \Phi - \bar \Phi$. Analogous definitions hold for the energy densities of the magnetic field, 
\begin{subequations}
\begin{align}
&
\overline{M}_p = \frac{1}{2} \langle \mid\nabla \times ( \nabla \bar h \times \vec r )
\mid^2 \rangle , \quad \overline{M}_t = \frac{1}{2} \langle \mid\nabla \bar g \times
\vec r \mid^2 \rangle, \\
&
\check{M}_p = \frac{1}{2} \langle \mid\nabla \times ( \nabla \check h \times \vec r )
\mid^2 \rangle , \quad \check{M}_t = \frac{1}{2} \langle \mid\nabla \check g \times
\vec r \mid^2 \rangle.
\end{align}
\end{subequations}
The total magnetic energy density $M$ 
averaged over the fluid shell is thus given by 
$M=\overline{M}_p+\overline{M}_t+\check{M}_p+\check{M}_t$. 

Figure 6 is instructive in that it demonstrates both, convection in
the presence and in the absence of its dynamo generated magnetic
field. As is evident from the right part of figure 6 with $M \sim 0$
relaxation oscillations occur in which convection sets in nearly
periodically for short episodes once the energy $\overline{E}_t$ of
the differential rotation has decayed to a sufficiently low
amplitude. But as soon as convection grows in amplitude, the
differential rotation grows as well and shears off the convection
columns. After convection has stopped the differential rotation decays
on the viscous time scale until the process repeats itself. 
\begin{figure}[t]
\psfrag{10000}{\hspace*{4mm}1}
\psfrag{20000}{\hspace*{4mm}2}
\psfrag{30000}{\hspace*{-4mm}$3\x10^5$}
\psfrag{0}{0}
\psfrag{1}{1}
\psfrag{5}{5}
\psfrag{3}{3}
\psfrag{0.00}{\hspace*{2mm}0.00}
\psfrag{0.25}{0.25}
\psfrag{0.50}{0.50}
\psfrag{t}{$t$}
\psfrag{M}{$M$}
\psfrag{Ex}{$M,E_x$}
\psfrag{Nu}{$Nu$}
%
%
\begin{center}
\vspace*{3mm}
\hspace*{-15mm} 
\epsfig{file=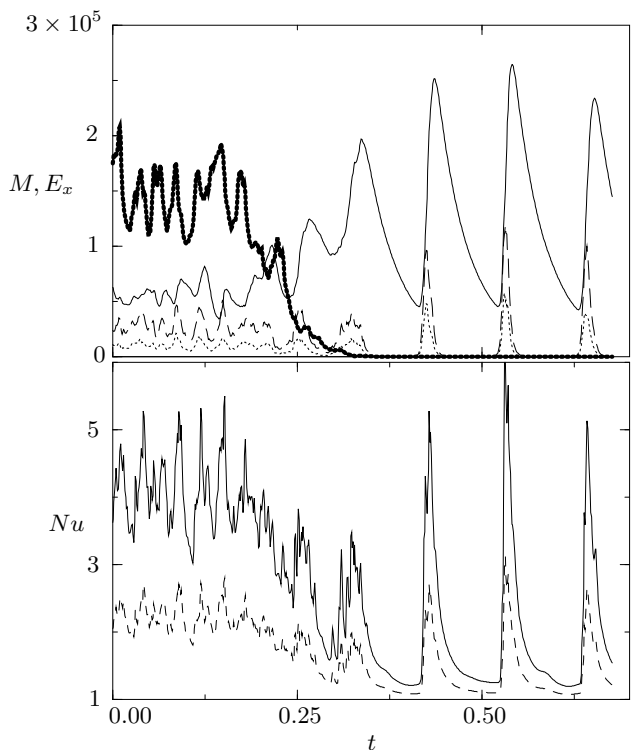,height=8.3cm,angle=-90}
\end{center}
\caption[]{Decay of a dynamo for  $P=0.5$, $\tau=1.5\x10^4$,
  $R=1.2\x10^6$,   $P_m=0.5$.  The total magnetic   energy density
  $M$ (thick dotted line, multiplied by a factor of 5), the  kinetic energy
  densities  $\overline{E}_t$ (solid line),
  $\check{E}_p$ (dotted line), $\check{E}_t$ (dashed line),  and
  Nusselt numbers  $Nu_i$ (solid line) at the inner and $Nu_o$ (dashed
  line) at the   outer spherical boundary (lower plot) are plotted as
  functions of   time $t$.}
\label{f6}
\end{figure}

As long as the magnetic field is present $\overline{E}_t$ is
suppressed owing to the action of the Lorentz force and high Nusselt
numbers are obtained. The dynamo generated magnetic field thus acts in
a fashion quite different from that of a homogeneous field which
typically counteracts the effects of rotation and tends to minimize
the critical value of the Rayleigh number when the Elsasser number,   
\begin{equation}
\Lambda= \frac{2 M P_m}{\tau}
\end{equation}
assumes the value $1$ in the case of a plane layer (Chandrasekhar
1961) or values of the same order in the case of a sphere (Fearn 1979)
or in the related annulus problem (Busse 1983) .  
\begin{figure}
\vspace*{4mm}
\npsfrag{0.1}{tl}{0.1}
\npsfrag{1.0}{tl}{1.0}
\npsfrag{Pm}{tl}{$P_m$}
\mypsfrag{R}{$R$}
\mypsfrag{aaa6}{\hspace{-1mm}$10^{6}$}
\mypsfrag{aaa7}{\hspace{-1mm}$10^7$}
%
\begin{center}
\hspace{-5mm}
\epsfig{file=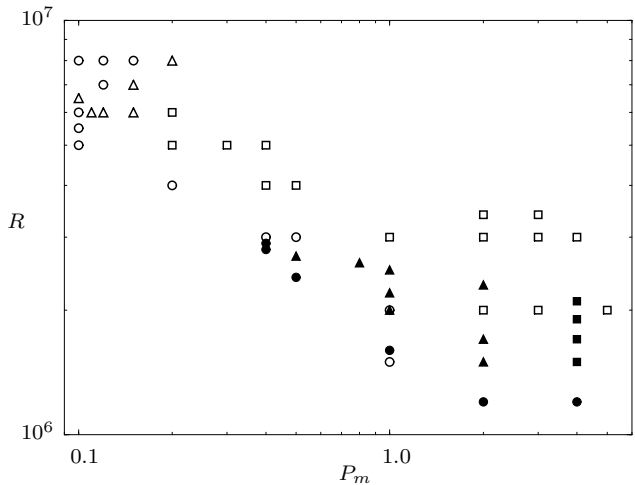,width=6cm,angle=-90}
\vspace*{2mm}
\end{center}
\caption[]{Convection driven dynamos as a function of the  Rayleigh
  number $R$ and the magnetic Prandtl number  $P_m$ for $P=0.1$,
  $\tau=10^5$ (empty symbols) and for $P=1$, $\tau=3\x10^4$ (filled
  symbols).
  The symbols indicate chaotic dipolar (squares), hemispherical
  (triangles), and decaying dynamos  (circles).}
\label{f.09}
\end{figure}

\section{5. Convection Driven Dynamos}
\label{sec:5}

It appears that dynamos are generated by convection in rotating
spherical shells for all parameter values  as long as the magnetic
Reynolds number is of the order $50$ or higher and the fluid is not
too turbulent. Since $Rm$ can be defined by $P_m \sqrt{2E}$ where the
kinetic energy density,
$E=\overline{E}_p+\overline{E}_t+\check{E}_p+\check{E}_t$, increases
with $R$, increasing values of the Rayleigh number are required for
dynamos as $P_m$ decreases. In planetary cores the latter parameter
may assume values as low as $10^{-5}$ or $10^{-6}$, but numerical
simulation have achieved so far only values somewhat below
$10^{-1}$. The trend of increasing $R$ with decreasing $P_m$ is
evident in figure 7 where results are shown for two different values
of $\tau$ and $P$. It is also evident from the results in the upper
left corner of the figure that an increasing $R$ may be detrimental
for dynamo action. This is a typical property of marginal dynamos at
the boundary of dynamos in the parameter space (Christensen \emph{et
  al.} 1999, Simitev and Busse 2005). Since convection at the relevant
values of $R$ is chaotic this property also holds for the generated
magnetic field. The magnetic energy must be finite near the boundary
of dynamos in the parameter space because a sufficient amplitude of
the fluctuating components of the convection flow can be obtained only
when the magnetic field is strong enough to suppress the differential
rotation. This property appears to hold even at high Prandtl numbers
where differential rotation occurs only in the form of a relatively
weak thermal wind.  
\begin{figure}[t]
\begin{center}
\hspace*{5mm}
\epsfig{file=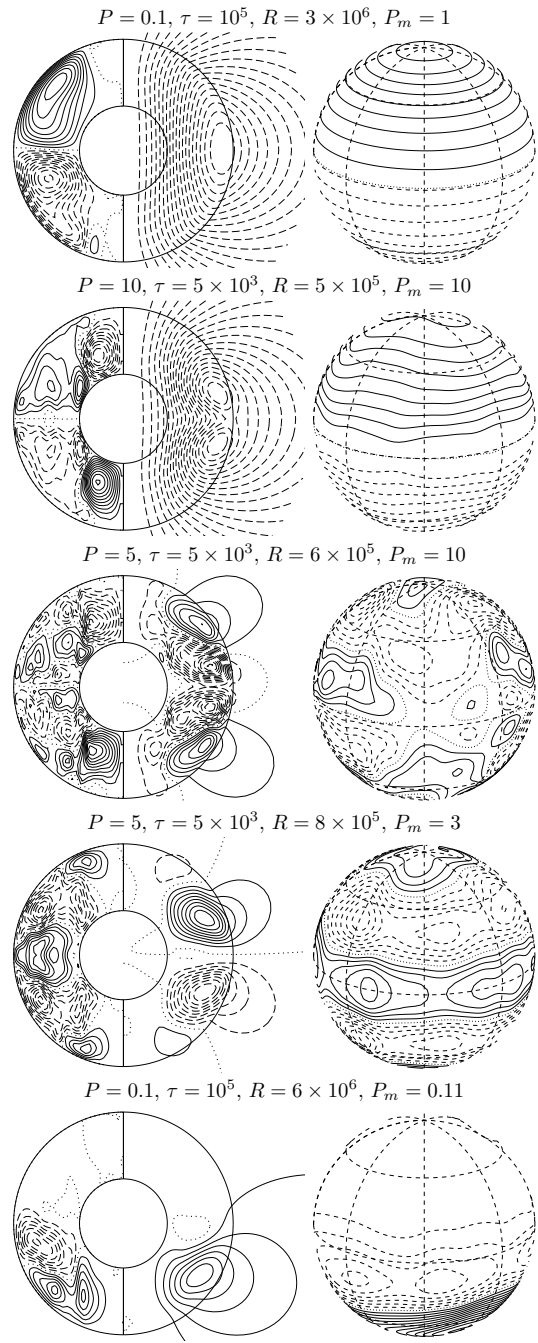,width=15cm}
\end{center}
\caption[]{Various types of dynamo symmetry.
  The left column shows lines of $\overline{B}_\varphi=$ const.\ (left
  half)   and  $r\sin\theta \dd_\theta\overline{h}=$ const.~(right
  half). The right column   shows surfaces of $B_r=$ const.~at
  $r=r_o+0.7$.} 
\label{f8}
\end{figure}

Even in its chaotic state convection at high values of $\tau$ exhibits
a strong symmetry with respect to the equatorial plane at high values
of $\tau$, at least as long as convection in the polar regions is
still weak. Because of this symmetry magnetic fields can be generated
either with the same symmetry as the convection flow, i.e. $h$ and $g$
are antisymmetric with respect to equatorial plane in which case one
speaks of a quadrupolar dynamo, or the magnetic field exhibits the
opposite symmetry with symmetric functions $h$ and $g$ in which case
one speaks of a dipolar dynamo. Of special interest are hemispherical
dynamos (Grote and Busse 2001) in which case the fields of dipolar and
quadrupolar symmetry have nearly the same amplitude such that they
cancel each other either in the Northern or the Southern
hemisphere. Examples of typical dynamos with different symmetries are
shown in figure 8. Here it is also evident that low $P$ and high $P$
dynamos differ in the structure of their mean toroidal fields. While
their mean poloidal dipolar fields exhibit hardly any difference, the
high $P$ dynamo is characterized by strong polar azimuthal flux tubes
which are missing in the low $P$ case. The reason for this difference
is that the radial gradient of the differential rotation in the polar
regions is much stronger for high $P$ than for low $P$. 

As the Rayleigh number increases and polar convection becomes stronger
the convection flow looses some of its equatorial symmetry and the
magnetic field can no longer easily be classified. Usually the dipolar
component becomes more dominant in cases of dynamos which started as
either quadrupolar or hemispherical dynamos at lower values of $R$.
\begin{figure*}
%
\begin{center}
\hspace*{-20mm}
\epsfig{file=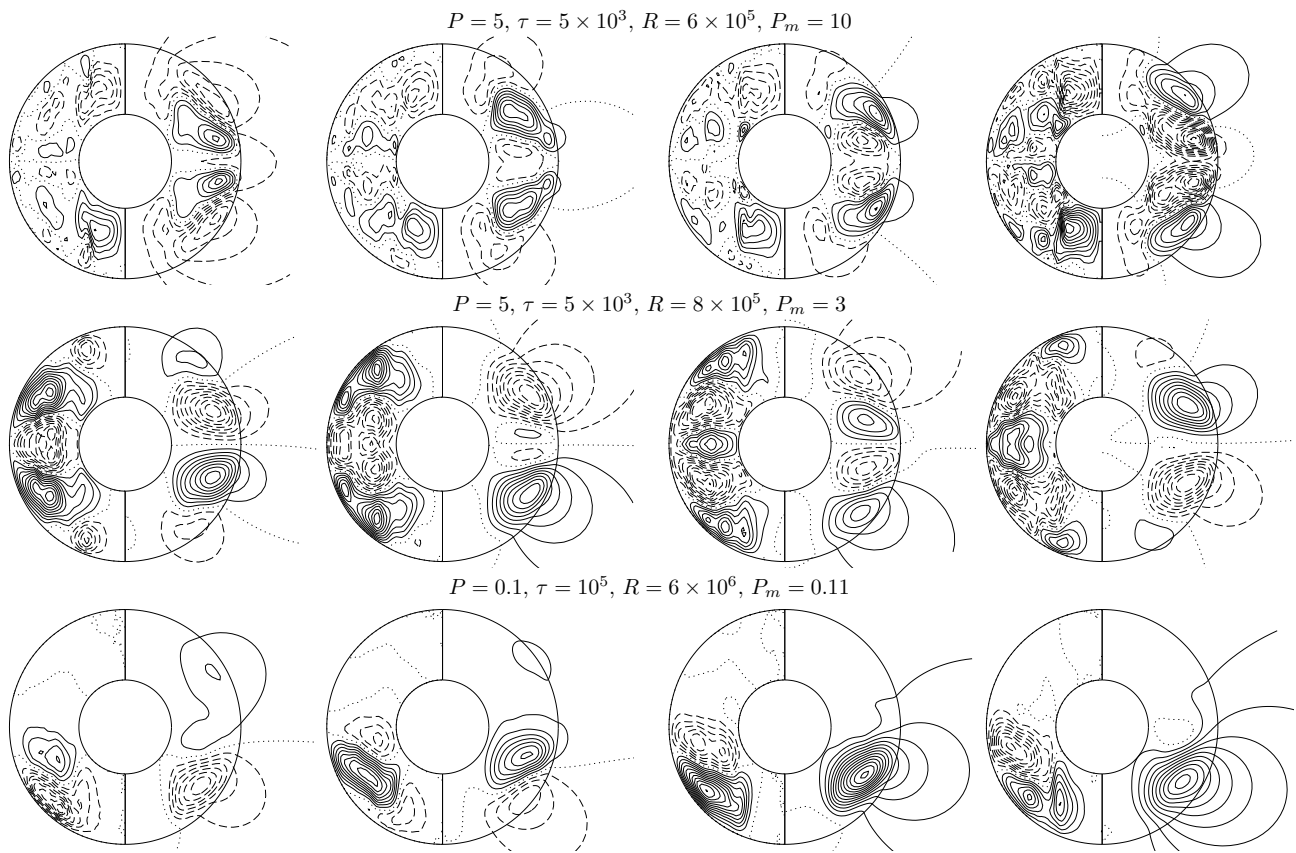,width=15cm}
\end{center}
\caption[]{Plots for the last three cases of figure 8 at the times 
  $n\cdot\Delta t$, $n=3,2,1,0$ (left to right) before the times of
  figure 8 with $\Delta t=$0.035,0.04,0.0025  (from top to bottom). }
\label{f9}
\end{figure*}
\begin{figure*}
\begin{center}
\epsfig{file=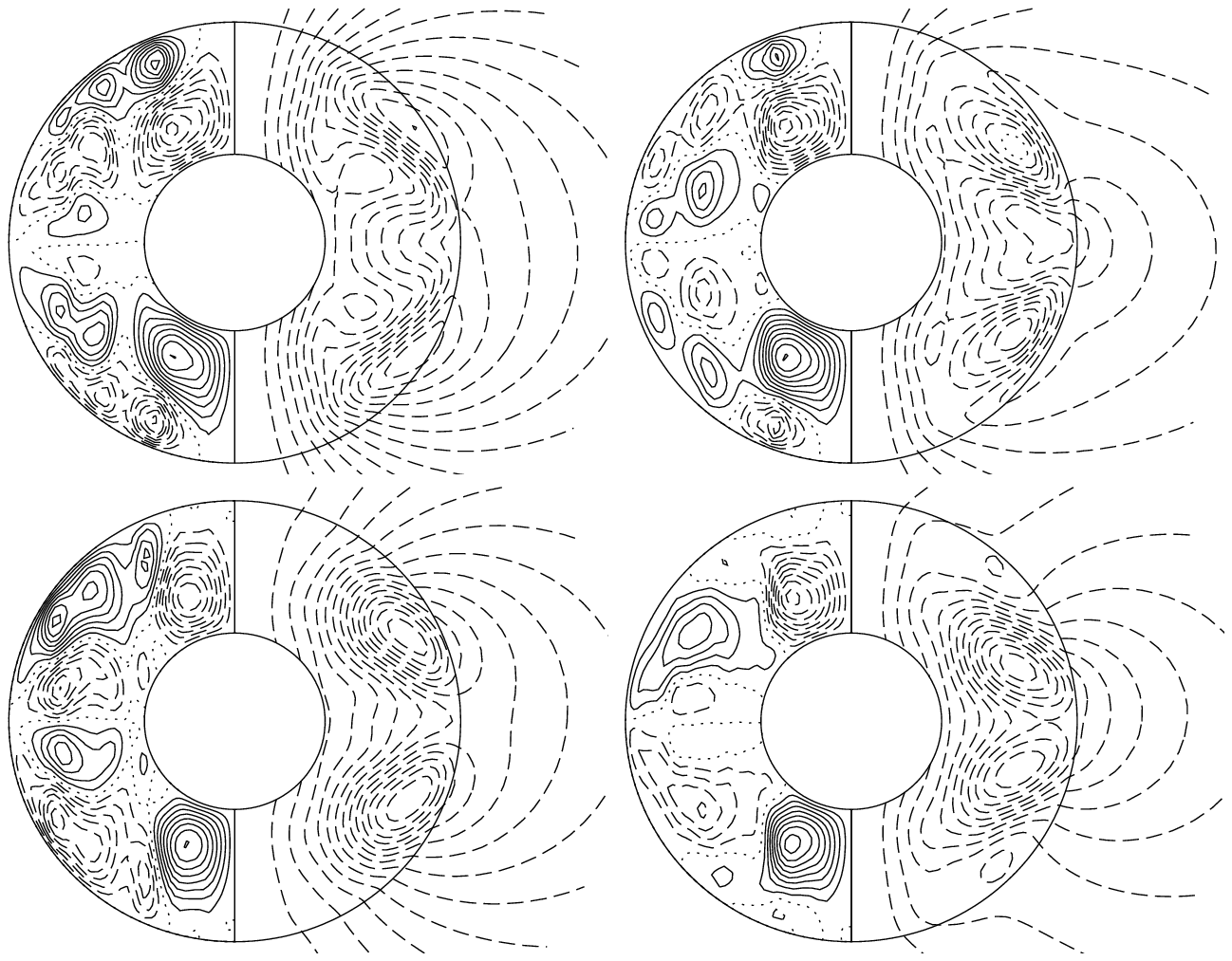,height=7cm}
\epsfig{file=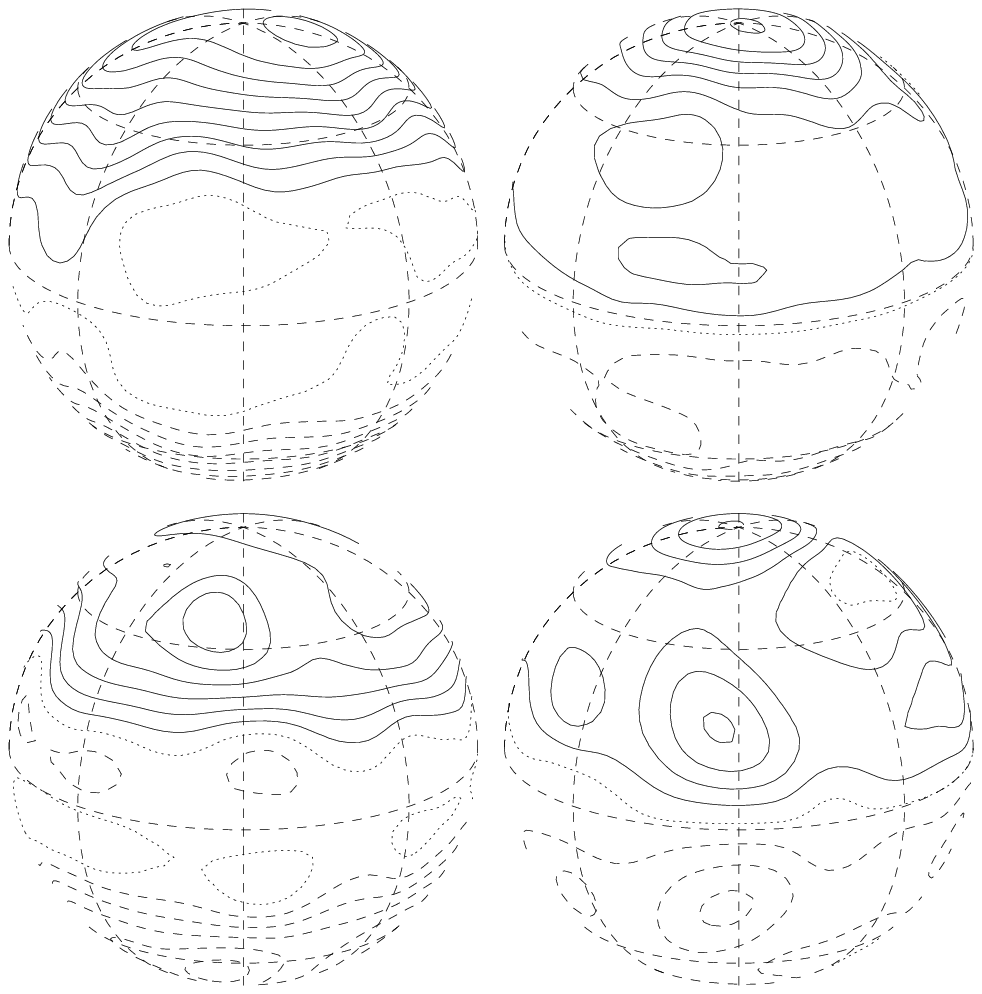,height=7cm}
\end{center}
\caption[]{An ``invisibly'' oscillating dynamo with $\Delta t = 0.04$
in  the case of $P=P_m=5$, $\tau=5 \x 10^3$ and $R=6 \x 10^5$.
The plots to the left show lines of constant
$\overline{B_{\varphi}}$
in their left halves and meridional field lines, $r \sin \theta
\dd_\theta \overline{h}  =$ const., in their right halves.
The plots to the right exhibits lines of constant $B_r$ on the surface
  $r=r_o+0.4$. The plots follow in clockwise sense  starting with the
  upper left one such that they complete approximately a full cycle.} 
\label{f10}
\end{figure*}
\begin{figure}
\epsfig{file=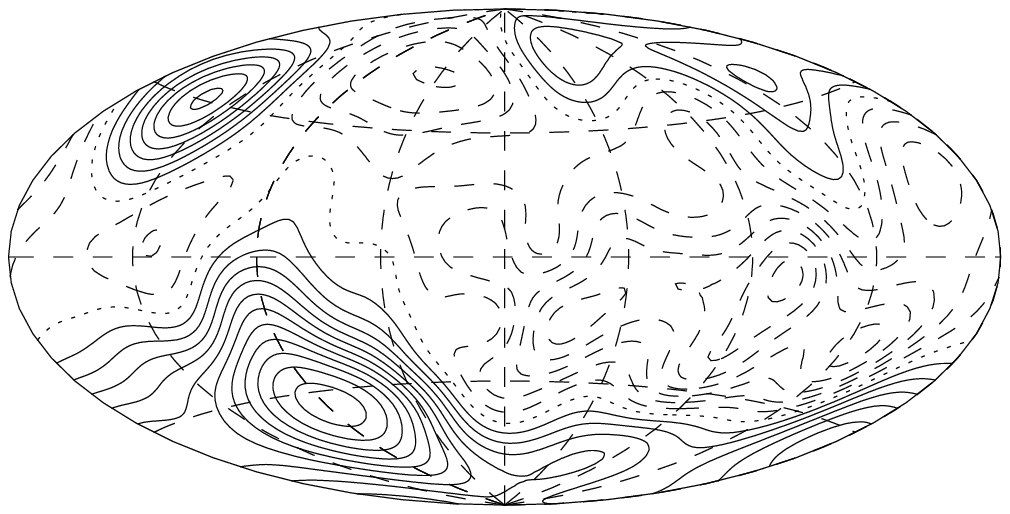,height=4cm,bbllx=147,bblly=464,bburx=442,bbury=615,clip=on}
\caption{
Lines of constant $B_r$ at $r=r_o+1$ for the dynamo with $R=2\times 10^6$, $\tau=5\times 10^3$, $P=P_m =1$.}
\end{figure}

It should be emphasized that quadrupolar and hemispherical dynamos are
usually oscillatory in that new azimuthal flux tubes of alternating
sign emerge in the equatorial plane, move towards higher latitudes,
while old flux tubes are dissipated in the polar region. There also
exist dipolar dynamos which exhibit the same type of
oscillations. They are typically found in the parameter space near the
region where hemispherical dynamos occur. To illustrate the dynamo
oscillations additional plots separated at equal distances in time
before the corresponding plot in the lower three rows of figure 8 are
shown in figure 9. The four plots in each case cover approximately
half a period of oscillation. It must be realized, of course, that the
oscillations are not strictly periodic since they occur in a turbulent
system. In this respect they resemble the solar cycle with its 22 year
period. Because the solar dynamo is believed to operate at the bottom
of the solar convection the propagation of the dynamo wave is towards
low latitudes on the sun. A remarkable feature of the dipole oscillation
in figure 9 is that the polar flux tubes hardly change in time. The
oscillation appears to be confined to the region outside the tangent
cylinder. At a lower value of $P_m$ the strong polar flux tubes even
inhibit the oscillation of the mean poloidal field as shown in figure
10. This case has been called the "invisibly" oscillating dynamo since
at some distance from the boundary of the spherical fluid shell the
oscillation of the dynamo can hardly be noticed.  
\begin{figure} \vspace*{4mm}
\mypsfrag{aa01}{$10^{-1}$}
\mypsfrag{aa0}{$10^0$}
\mypsfrag{aa1}{$10^1$}
\mypsfrag{EM}{\hspace{-1mm}$E_x$, $M_x$}
\mypsfrag{1aa0}{\hspace{2mm}$10^0$}
\mypsfrag{1aa1}{\hspace{2mm}$10^1$}
\mypsfrag{1aa2}{\hspace{2mm}$10^2$}
\mypsfrag{1aa3}{\hspace{2mm}$10^3$}
\mypsfrag{1aa4}{\hspace{2mm}$10^4$}
\mypsfrag{1aa5}{\hspace{2mm}$10^5$}
\mypsfrag{1aa6}{\hspace{2mm}$10^6$}
\mypsfrag{2aa1}{\hspace{2mm}$10^1$}
\mypsfrag{2aa2}{\hspace{2mm}$10^2$}
\mypsfrag{2aa3}{\hspace{2mm}$10^3$}
\mypsfrag{2aa4}{}
\mypsfrag{2aa5}{\hspace{2mm}$10^5$}
\mypsfrag{2aa6}{\hspace{2mm}$10^6$}
\mypsfrag{7.5}{7.5}
\mypsfrag{10} {10}
\mypsfrag{3}{3}
\mypsfrag{2}{2}
\mypsfrag{5}{5}
\mypsfrag{1} {1}
\mypsfrag{0} {0}
\mypsfrag{c2}{\hspace{-0mm}$P=1$}
\mypsfrag{f1}{\hspace{-6mm}$P=3$}
\mypsfrag{f3}{\hspace{-6mm}$P=5$}
\mypsfrag{f5}{\hspace{-6mm}$P=15$}
\mypsfrag{d} {\hspace{-0mm}$P=1$}
\mypsfrag{g1}{\hspace{-4mm}$P=2$}
\mypsfrag{g3}{\hspace{-4mm}$P=5$}
\mypsfrag{g5}{\hspace{-5mm}$P=15$}
\mypsfrag{VO}{\hspace{-1mm}$V_x$, $O_x$}
\mypsfrag{R}{}
\mypsfrag{RR}{\hspace{0mm}$P_m$}
\begin{center}
\hspace{10mm}
\hspace*{-12mm}
\epsfig{file=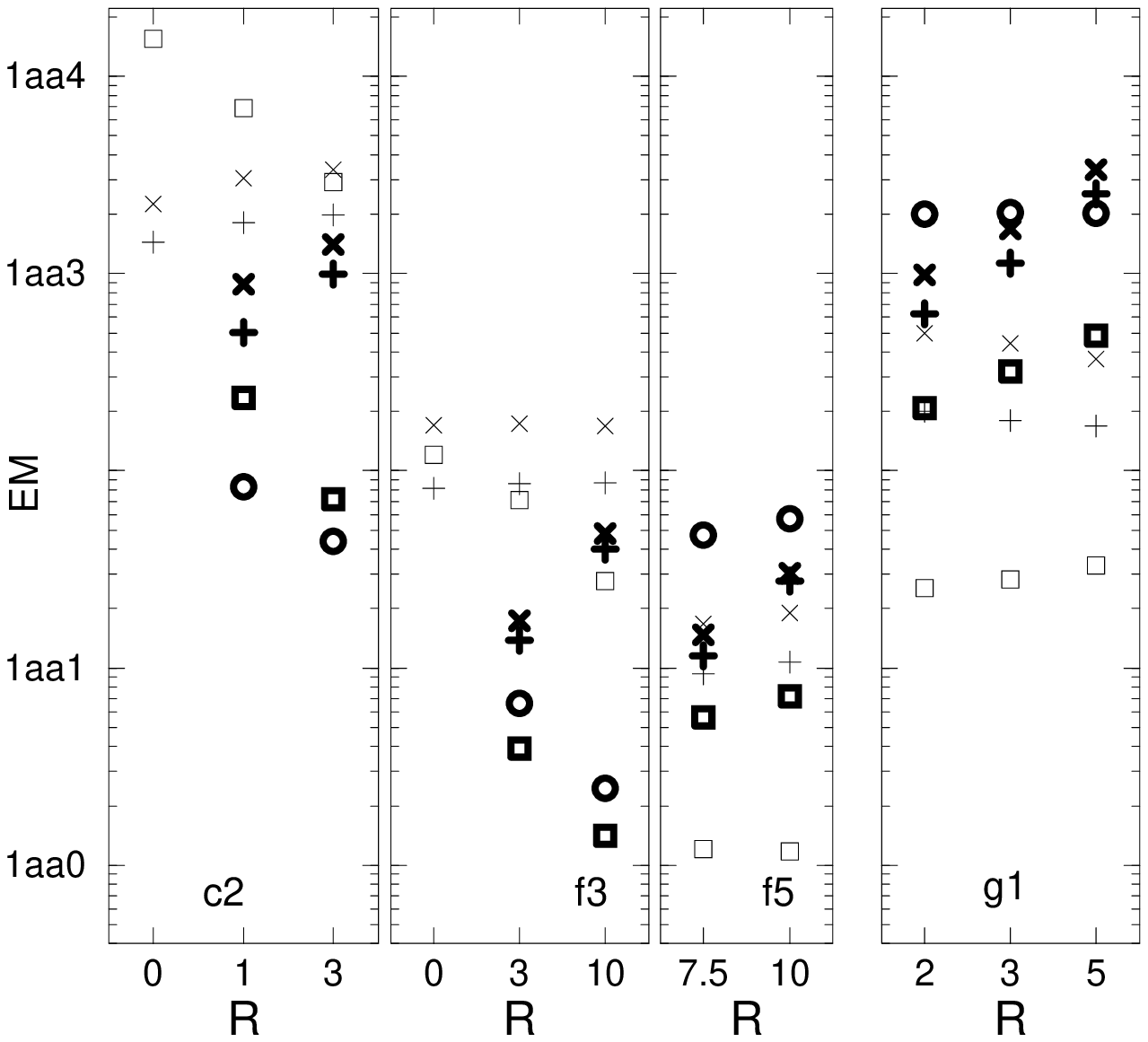,width=7.2cm,height=8.5cm,angle=-90}\\
\epsfig{file=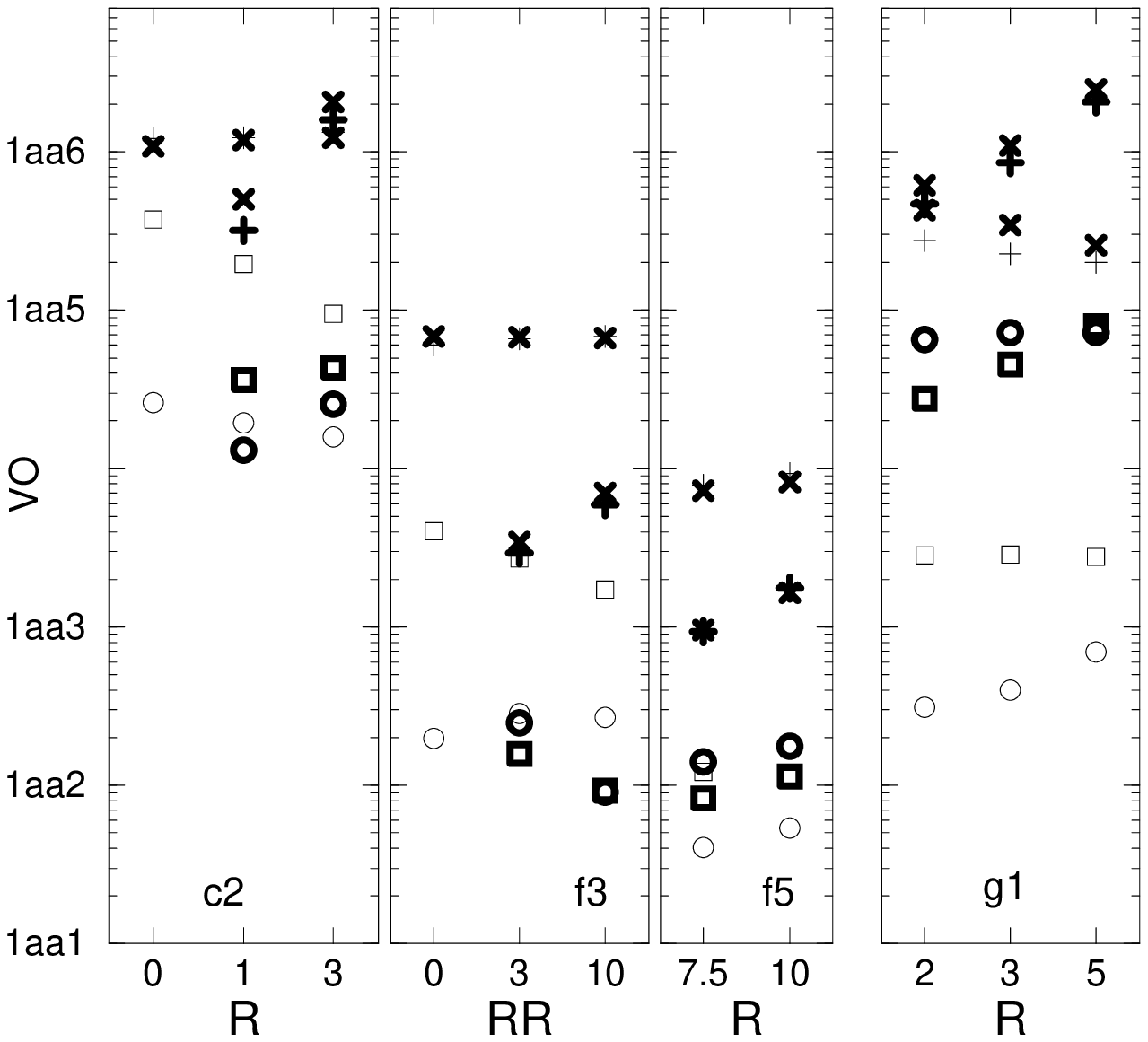,width=8cm,height=8.5cm,angle=-90}
\end{center}
\caption[]{Kinetic $E_x$ and magnetic $M_x$ energy densities (upper part)
and viscous $V_x$ and Ohmic $O_x$ dissipations (lower part) as
functions of $P_m$ for convection driven dynamos
for $\tau=5\x10^3$, $R=10^6$ (left three columns), $\tau=3\x10^4$,
$R=3.5\times10^6$ (right column) and Prandtl  number
as indicated in the boxes. The highest values of the Elsasser number
for the cases $P=$ 1, 5, 15, 2 are  $\Lambda=3.02$, $0.37$,
$0.49$ and $1.12$ respectively. The
components $\overline{X}_p$, $\overline{X}_t$,
$\check{X}_p$, $\check{X}_t$ (where $X = E$, $M$, $V$, $O$) are
represented by circles, squares, plus-signs and
crosses, respectively. Kinetic energy densities and viscous
dissipations are shown with light symbols, magnetic energy
densities and Ohmic dissipations are shown with heavy symbols. The
magnetic Prandtl numbers $P_m=0$ correspond to non-magnetic
convection.}
\label{f11}
\end{figure}

The relatively simple magnetic fields displayed in figures 8, 9, and
10 should not be regarded as representative for high values of the
Rayleigh number when $R$ exceeds its critical value $R_c$ by an order
of magnitude or more. In that case the equatorial symmetry has nearly
disappeared owing to the dominant convection in the polar regions and
non-axisymmetric components of the field tend to exceed axisymmetric
ones as shown , for example, in figure 11. The strong growth with
increasing $Rm$ of the non-axisymmetric magnetic flux is also evident
in figure 12 when energies corresponding to different values of $P_m$ are compared. 

An important question is the average strength of the magnetic field
generated by the dynamo process in dependence on the parameters. Two
concepts are often used in attempts to answer this question. In
astrophysical situations the equipartition between magnetic and
kinetic energy is used as a guide. 
While such a balance may hold locally as, for example, in the
neighborhood of sunspots, it is not likely to be applicable to global
planetary magnetic fields. In the Earth's core, for example, the
magnetic energy density exceeds the kinetic one by a factor of the
order of $10^3$. The second concept is based on the property that an
Elsasser number $\Lambda$ of the order unity appears to correspond to
optimal  conditions for convection in rotating systems in the presence
of an applied nearly uniform magnetic field. While this idea may be
useful as a first rough estimate, the results of numerical dynamo
simulations do not support this concept very well. The examples shown
in figure 12 exhibit $\Lambda$-values differing by an order of
magnitude and even larger variations have been reported by Simitev and
Busse (2005) and by Christensen and Aubert (2005).  

Another important feature demonstrated in figure 12 is the change in
the structure of the magnetic field with increasing Prandtl
number. While for $P=5$ and lower values the mean poloidal field is
small in comparison with the fluctuating components, this situation
changes dramatically at about $P=8$ for $\tau=5\cdot10^3$ such that
for higher $P$ the mean poloidal field becomes dominant. Usually this
field is dipolar. This change is associated with the transition from
the geostrophic differential rotation generated by Reynolds stresses
to the thermal wind type differential rotation caused by a latitudinal
temperature gradient. While the magnetic energy $M$ may exceed the total kinetic energy $E$ by orders of magnitude in particular for high Prandtl numbers, ohmic dissipation is usually found to be at most comparable to viscous dissipation in numerical simulations. But this may be due to the limited numerically accessible parameter space.

The brief introduction of this section to convection driven dynamos in
rotating spherical fluid shells can only give an vague impression of
the potential of numerical simulations for the understanding of
planetary magnetism. Many more examples of such simulations can be
found in the literature. Usually they have been motivated by
applications to the geodynamo and $P=1$ is assumed in most cases for
simplicity. For some recent systematic investigations we refer to
Christensen et al. (1999), Grote et al. (2000), Jones (2000), Grote
and Busse (2001), Kutzner and Christensen (2000, 2002), Busse et
al. (2003), Simitev and Busse (2005) and other papers referred to
therein. 

\section{6. Applications to Planetary Dynamos}
\label{sec:6}

\subsection{6.1 General Considerations}

In many respects it is too early to model the dynamo process in
particular planets. The numerical simulations of the kind discussed in
the preceding sections are still rather removed from the parameter
regime relevant to planetary interiors. Only most recently attempts
have been made to extrapolate results to high Rayleigh numbers and
high values of $\tau$ (Christensen and Aubert, 2006). Moreover, only
the most important physical parameters have been taken into account
and relevant properties such as compressibility and other deviations
from the Boussinesq approximation have been considered only in special
models applied to the Earth's core (Glatzmaier and Roberts 1996) or to
the sun (Brun \emph{et al.} 2004). On the other hand, many essential
parameter values of planetary cores are not sufficiently well  known
to provide a basis for the development of specific dynamo models. Much
of the future progress of the field will thus depend on the mutual
constraints derived from observational evidence and from theoretical
conclusions in order to arrive at a better understanding of the
workings of planetary dynamos. 
\begin{table*}
\begin{tabular}{l@{\extracolsep{5mm}}c@{\extracolsep{5mm}}c@{\extracolsep{5mm}}c@{\extracolsep{5mm}}c@{\extracolsep{5mm}}c}\hline
\rule{0mm}{4mm}Planet      & Planetary Radius& Dipole Moment & Core Radius & Angular
            Rate of& Magn. Diffus. \\
(Satellite) &  $R_p$ ($10^6$ m) & ($10^{-4}$T$\times R_p^3$) &($10^6$ m)
          & Rot. $\Omega$ ($10^{-5}$s$^{1}$) & (m$^2$/s)\\[2mm] \hline
\rule{0mm}{4mm}Mercury & 2.439 & 0.0025 & 1.9 & 0.124 & $\sim$ 2 \\
Earth   & 6.371 & 0.31   & 3.48 & 7.29 & 2 \\
Mars    & 3.389 & $3^\dagger$ & $\sim$ 1.5 & 7.08 &$\sim$ 2\\
Jupiter & 69.95 & 4.3 & $56^\ast$  & 17.6 & $30^\ast$ \\
Saturn  & 58.30 & 0.21 & $29^\ast$ & 16.2 & $4^\ast$ \\
Uranus  & 25.36 & 0.23 & $18^\ast$ & 10.1 & $100^\ast$ \\
Neptune & 24.62 & 0.14 & $20^\ast$ & 10.8 & $100^\ast$ \\ 
Ganymede& 2.63 & 0.0072& 0.7 & 1.02 & 4 \\[2mm]\hline
\end{tabular}\\[2mm]
$^\ast$ Since the Giant Planets do not possess a
  well-defined boundary of a highly conducting core, \\ values of the
  most likely region of dynamo activity are given. \\
$^\dagger$ Remnant magnetism of the Martian crust appears to require
  an ancient dynamo \\ with a field strength of at least 10 times the
  Earth's magnetic field.
\caption[]{Planetary parameter values (after Stevenson (2003), Nellis
  \emph{et al.} (1988), Connerney (1993) and various websites).}
\end{table*}

In table 2 some properties related to planetary magnetism have been listed. The dipole moments of the planets have been given as multiples of $G\cdot R_p^3$ where $R_p$ denotes the mean radius of the planet or satellite. Thus the numerical value indicates the field strength in $Gauss$ in the equatorial plane of the dipole at the distance $R_p$ from its center. We have included Mars in the table, although it does have an active dynamo at the present time. The strong magnetization of the Martian crust indicates, however, that a strong field dynamo must have operated in the early history of the planet.

The most important question of the theory of planetary magnetism is
the dependence of the observed field strength on the properties of the
planet. We have already discussed in the historical introduction
various proposals for such dependences based on {\it ad hoc}
assumptions, but these have been abandoned by and large. Only the
concept of an Elsasser number $\Lambda$ of the order unity is still
frequently used. Stevenson (2003) shows that a value $\Lambda = 0.3$
fits most of the planets with a global magnetic field quite well, but
in cases such as Mercury and Saturn only a value of the order
$10^{-2}$ can be estimated for $\Lambda$. In order to save the concept
different dynamo regimes must be assumed. Equilibration balances for
weaker magnetic fields have been proposed as, for instance, in the
case when the strength of convection is not sufficient to attain the
$\Lambda \sim 1$ balance. For details see Stevenson (1984). 

An important balance often invoked in discussions of planetary dynamos
(see, for example, Stevenson (1979) and Jones (2003)) is the
MAC-balance (Braginsky 1967) where it is assumed that Lorentz force,
buoyancy force and pressure gradient are all of the same order as the
Coriolis force, while viscous friction and the momentum advection term
are regarded as negligible in the equations of motion. The neglection
of the latter term is well justified for high Prandtl numbers, but it
is doubtful for values of $P$ of the order unity or less. Inspite of
their smallness in comparison with the Coriolis force, the divergence
of the Reynolds stress can generate the most easily excitable mode of
a rotating fluid, namely the geostrophic differential rotation which
can not be driven by the Coriolis force, the pressure gradient or
buoyancy. Of course, an excitation by the Lorentz force is possible in
principle. The latter force, however, usually inhibits the geostrophic
differential rotation. Since the mean azimuthal component of the
magnetic field is typically created through the shear of the
differential rotation, the  Lorentz force opposes the latter according
to Lenz' rule. 

Christensen and Aubert (2006) have recently introduced a concept in
which the equilibrium strength of planetary magnetism  does no longer
depend on the rate of rotation $\Omega$, but instead depends only on
the available power for driving the dynamo. They find that scaling
laws can be obtained once Rayleigh number, Nusselt number and
dimensionless buoyancy flux $\hat Q$ have been defined through
quantities that no longer involve molecular diffusivities. This is a
surprising result since it requires the presence of rotation and can
not be achieved in a non-rotating system.  The final estimate obtained
for the strength of the magnetic field in the dynamo region fits the
cases of the Earth and Jupiter quite well, but yields the result that
Mercury's field could not be generated by a buoyancy driven dynamo
since the magnetic Reynolds number would be too small. In the cases of
the outer planets relevant parameters are not sufficiently well known
to draw definitive conclusions. 
\begin{figure}
\epsfig{file=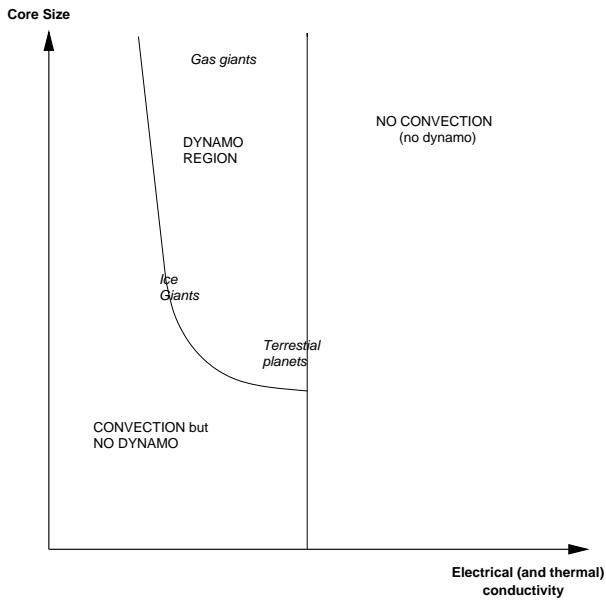,width=8cm}
\caption{Sketch of Stevenson's paradox (after Stevenson 2003).}
\end{figure}

Finally we like to draw attention to what has been called Stevenson's
paradox, namely that a too high electrical conductivity is detrimental
for dynamo action in planetary cores. According to the Wiedemann-Franz
law of condensed matter physics the thermal conductivity $k$ of a
metal is proportional to its electrical conductivity times the
temperature, $k = \sigma T L$, where L has approximately the same
value for all metals. This is due to the fact that in metals electrons
dominate both, the transports of heat and of charge. As a consequence
there is an upper bound on the electrical conductivity for which a
dynamo driven by thermal convection is possible because a high value
of $\sigma$ implies a low temperature change with increasing depth
needed to carry the heat from the interior of the planetary core. But
as soon as the temperature change with pressure falls below its
isentropic value thermal convection disappears and a stably stratified
environment is obtained which makes it difficult to drive a dynamo by
other types of motions such as compositional convection. Ignoring the
latter possibility Stevenson (2003) has sketched the diagram shown in
figure 13 where the marginal nature of dynamos in terrestrial as well
as in the "icy" planets is indicated. 

In the following brief characterizations of the magnetisms of various
planets and moons are given. In terrestrial planets and satellites
dynamos are possible in the liquid parts of their iron cores. Such
cores always include light elements which depress the temperature of
freezing and can give rise to compositional buoyancy in the presence
of a solidifying inner part of the core. In Jupiter and especially in
Saturn convection may also be driven in part by compositional buoyancy
since hydrogen and helium are immiscible in certain regions and helium
drops may rain out. In numerical dynamo models the different sources
of buoyancy are usually not distinguished, however.   

\subsection{6.2 Mercury}
 
It has already become apparent that the origin of Mercury's magnetism
 is especially enigmatic. Besides the hypothesis of an active dynamo,
 thermoelectric currents and remnant magnetism of the crust compete
 as explanations. The latter possibility had been discarded for a long
 time since Runcorn's theorem states that a homogeneous spherical
 shell can not be magnetized by an interior magnetic field in such a
 way that a dipolar field can be observed from the outside after the
 interior source has been removed (Runcorn, 1975). Aharonson et
 al. (2004) have shown, however, that plausible  inhomogeneities of
 the crust could possibly explain a remnant magnetic field created by
 an ancient dynamo. 
 
The idea that a convection driven dynamo operates in a fluid outer
core of Mercury has received some support from observations of
Mercury's librations (Margot et al., 2004) and from numerical
simulations (Stanley et al., 2005). The amplitudes of the small
periodic variations (librations) of Mercury's rotation caused by the
gravitational pull of the sun on the non-axisymmetric mass
distribution of Mercury appear to be too large for a fully solidified
planet. This suggests that the solid inner core is separated from the
mantle by a fluid outer core in which convection flows may occur. The
numerical simulations of dynamos in the thin shell of the outer core
have demonstrated that the dipole strength measured from the outside
may not be representative for the strength of the magnetic field at
the core-mantle boundary and the criteria mentioned above may be
satisfied after all. The question, however, whether there is a
sufficiently strong source of buoyancy to sustain convection and the
dynamo over the age of the planet has not yet been answered
satisfactorily. 
 
 \subsection{6.3 Venus}
 
 Space probes have clearly shown that there is no dynamo operating
 presently in Venus. This result is surprising since Venus is very
 similar to the Earth in many other aspects. That the period of
 rotation is only $1/243$ of a day should not matter much since the
 Coriolis parameter $\tau$ is still huge.  More important is the fact that
 the cooling of the planet is less efficient because of the apparent
 absence of mantle convection and plate tectonics. A solid inner core
 may not have yet started to grow in Venus and vigorous convection
 does not occur in the core, at least not at the present time. 
 
 \subsection{6.4 Mars and Moon}
     
Neither Moon nor Mars have an active dynamo, but the magnetized rocks
in their crusts suggest that  in their early history about 4 billion
years ago dynamos may have been operating in the iron cores of these
bodies. The small lunar core with a radius of the order of 350 km and
the age of the magnetized lunar rocks put severe constraints on a
possible dynamo origin of lunar magnetism as discussed by Stegman et
al. (2003). In their model these authors try to accommodate in
particular the apparent sudden onset of a lunar dynamo at a time
$3.5\, Ga$ ago. Because of its larger iron core the possibility of a
temporarily operating Martian dynamo is more likely, but the strong
magnetization of parts of the Martian crust requires a field of at
least ten times the strength of the present geomagnetic field. For a
recent review of the implications of Martian magnetism for the
evolution of the planet we refer to Stevenson (2001). 

\subsection{6.5 Jupiter}

The Jovian magnetic field is rather similar to the geomagnetic one
sharing with it a dipole axis that is inclined by an angle of the
order of 10 degrees with respect to the axis of rotation. That the
higher harmonics of the field are relatively stronger than those in
the case of the Earth suggest that the Jovian dynamo is driven at a more shallow
layer than the Earth's outer core. This assumption is in accordance with the
transition to metallic hydrogen which is expected to occur at a
pressure of $140 GPa$ corresponding to a region with a radius of 0.9
of the Jovian radius (Nellis et al., 1996). The transition is not a
sharp one as suggested by earlier models of Jupiter's interior, but a
gradual one leading to a decrease of the magnetic diffusivity down to
about $4m^2/s$. Since this value yields a rather high value for the
Elsasser number $\Lambda$ Stevenson (2003) has argued that the dynamo
is most active above the region of highest conductivity. Jones (2003),
on the other hand, accepts the value $20$ for $\Lambda$, but must
admit a high magnetic Reynolds number of the order $10^4$ at which an
effective dynamo may no longer be possible. A detailed model of a
convection driven Jovian dynamo should include effects of
compressibility and depth dependent electrical conductivity, but none
has yet been published. 

\subsection{6.6 Saturn}

Saturn's magnetic field with a strength of $0.2mT$ near the poles is
much weaker than that of Jupiter which must be attributed primarily to
its deeper origin in the planet. The transition pressure of $140GPa$
is reached only at about half the planetary radius. A property that
has received much attention is that the Saturnian magnetic field is
nearly axisymmetric with respect to the axis of rotation. This
property can not be interpreted as a contradiction to Cowling's (1934)
theorem since it has long been shown that fields with arbitrarily small
deviations from axisymmetry can be generated by the dynamo process
(Braginsky 1976). Stevenson (1982) gave a reasonable explanation for
the almost axisymmetric Saturnian field by demonstrating that the
differential rotation in the stably stratified, but still electrically
conducting gas shells above the dynamo region will tend to shear off
all non-axisymmetric components of the field. The problem may be a bit
more intricate as has been pointed out by Love (2000) who showed that
sometimes fields with solely non-axisymmetric components are found in
such situations.

\subsection{6.7 Uranus and Neptune} 

A metallic liquid probably does not exist in Uranus and Neptune whose
interiors consist mostly of ice and rocks. Because of the pressure
dissociation of water deeper regions of these planets are
characterized by an ionic conductivity which is lower than typical
metallic conductivities by between one and two orders of
magnitude. The mixture of water, methane, ammonia and other ices is
sufficiently fluid that convection can occur and that a dynamo is thus
possible. The fact that the observed magnetic field do not show an
alignment with the axis planetary rotation and that quadrupolar and
octupolar components are comparable to the dipolar components  of the
fields has been attributed to a dynamo operating in a thin shell
(Ruzmaikin and Starchenko, 1991; Stanley and Bloxham, 2004). This seems be hardly necessary, however,
since even convection driven dynamos in thick shells often exhibit
such fields when small scale components dominate as in the case of
high Rayleigh numbers and low Prandtl numbers. See, for example,
figure 11 which shows a magnetic field rather similar to that of
Uranus as displayed in the paper of Connerney (1993). A special
difficulty for a convection driven dynamo in the interior of Uranus
is caused by its low emission of heat. Holme and Bloxham (1996)
suggest that a typical dynamo would involve more Ohmic dissipation
than corresponds to the heat flux emitted from the interior of the
planet.

\subsection{6.8 Ganymede and other Satellites}

It came as a great surprise when the measurements of the Galileo
spacecraft indicated that Jupiter's moon Ganymede possesses a global
magnetic field for which an active dynamo inside the satellite seems
to be the only realistic explanation (Kivelson et al., 1996), although
the possibility of a remnant magnetism can not easily be excluded
(Crary and Bagenal 1998). It must be kept in mind that Ganymede is the
largest satellite in the solar system which exceeds even Mercury in
size. Nevertheless the estimated radius of its iron core is only about
660 km and it is hard believe that it could still be partly molten
unless it contain a lot of radioactive potassium 40 or Ganymede was
captured into a resonance in its more recent history (Showman \emph{et
  al.} 1997). The fact that Ganymede's magnetic moment is nearly
aligned with the ambient Jovian magnetic has led to the suggestion
that the ambient field could aid significantly Ganymede's
dynamo. Sarson et al. (1997) have investigated this question with the
result that the ambient field is too weak to exert much influence. On
the other hand, in the case of Io which is much closer to Jupiter the
interaction between the ambient field and the liquid iron core can
explain the observed structure of the magnetic field without the
assumption of an active dynamo. 

There are no other satellites in the solar system where an active
dynamo must be suspected. The substantial magnetization of many iron
meteorites suggest, however,  that several differentiated
proto-planets have had dynamos in the early days of the solar system. 
     
\section{7. Concluding Remarks}
\label{sec:7}
It is apparent from the above discussions that dynamo theory does not
yet have much specific information to contribute to the interpretation
of the observed magnetic properties of planets and satellites in the
solar system. Even possible interactions between thermal and
compositional buoyancies have not yet been taken fully into account
(Glatzmaier and Roberts 1996, Busse 2002b). Some typical properties
are already apparent, however, as for instance: 
\begin{itemize}
\item Dynamos that are dominated by a nearly axial dipole and exhibit
  a magnetic energy that exceeds the kinetic one by orders of
  magnitude as in the case of the Earth are typical for high effective
  Prandtl numbers as must be expected when convection is primarily
  driven by compositional buoyancy. The rather Earth like appearance
  of the dynamo of Glatzmaier and Roberts (1995) is in part due to the
  high value of $P$ used in their computational model. 
\item Dynamos exhibiting strong higher harmonics are more likely
  driven by thermal convection corresponding to Prandtl numbers of the
  order unity or less. 
\item Dynamo oscillations are a likely phenomenon in the presence of a
  sufficiently strong differential rotation. They may not always be
  apparent in the poloidal magnetic field seen at a distance from the
  dynamo region. 
\item Considerations based on the effects on convection of an imposed
  homogeneous magnetic field 
can not directly be applied to the case of convection driven dynamos.
\end{itemize}
More details on the parameter dependence of dynamos will certainly
emerge in the future as the increasingly available computer capacity
will allow extensions of the parameter space accessible to computer
simulations. Space probes such as MESSENGER in the case of Mercury
will provide much needed detailed information on planetary magnetic
fields. Eventually we may learn more about their variation in time
which is one of the most interesting properties of planetary magnetic
fields. 

As in the case of stellar magnetism where the study of star spots and
stellar magnetic cycles is contributing to the understanding of solar
magnetism it may eventually become possible to learn about the
magnetism of extrasolar planets and apply this knowledge for an
improved understanding of solar system dynamos. Undoubtedly the field
of planetary magnetism will continue to be an exciting one!  


\section*{8. References}
\begin{description}
\item[]
Aharonson O, Zuber M T, Solomon S C 2004 Crustal remanence in an internally magnetized non-uniform shell: a possible source for Mercury's magnetic field? {\it Earth Plan. Sci. Lett.} {\bf 218} 261-268
\item[]
Ardes M, Busse F H, Wicht J 1997 Thermal convection in rotating
spherical shells. {\it Phys. Earth Plan. Int.}, {\bf 99}, 55-67
\item[]
 Backus G 1958 A class of Self-Sustaining Dissipative Spherical Dynamos. {\em Annals Phys.} 4, 372-447
\item[]
Braginsky S I 1967 Magnetic waves in the earth's core. {\it Geomagn. Aeron.}, {\bf 7}, 851-859
\item[]
Braginsky S I 1976 On the nearly axially-symmetrical model of the hydromagnetic dynamo of the Earth. {\it Phys. Earth Plan. Int.}, {\bf 11}, 191-199
\item[]
Brun A S, Miesch M S, Toomre J 2004 Global-Scale Turbulent Convection and Magnetic Dynamo Action in the Solar Envelope. {\it Astrophys. J.} {\bf 614}, 1073-1098
\item[]
Bullard E C 1949 The magnetic field within the earth. {\it Proc. Roy. Soc. London}, {\bf A 197}, 433-463   
\item[]
Burke B F, Franklin K L 1955 Observations of variable radio source associated with the planet Jupiter. {\it J. Geophys. Res.}, {\bf 60}, 213-217
\item[]
Busse F H 1976 Generation of Planetary Magnetism by Convection. {\it Phys. Earth Plan. Inter.}, {\bf 12}, 350-358
\item[]
Busse F H 1983 Recent developments in the dynamo theory of planetary magnetism,
{\it Ann. Rev. Earth Plan. Sci.} {\bf 11}, 241-268
\item[]
Busse F H 2002a Convection  flows in rapidly rotating spheres and their
 dynamo action. {\it Phys. Fluids}, {\bf 14}, 1301-1314 
\item[]
Busse F H 2002b Is low Rayleigh number convection possible in the Earth's
core? {\it Geophys. Res. Letts.}, {\bf 29}, GLO149597 
\item[]
Busse F H, Carrigan C R 1976 Laboratory simulation of thermal
convection in rotating planets and stars, {\it SCIENCE}, {\bf 191},
81-83 
\item[]
Busse F H,  Cuong P G 1977 Convection in rapidly rotating spherical
fluid shells. {\it Geophys. Astrophys. Fluid Dyn.} {\bf 8}, 17--41
\item[]
Busse F H, Grote E, Simitev R 2003 
  Convection in rotating spherical shells and its dynamo
    action, pp. 130-152 {\it  in "Earth's Core and Lower Mantle", C.A.Jones,
  A.M.Soward and K.Zhang, eds., Taylor \& Francis}
\item[]
Busse F H, Heikes K E 1980 
Convection in a Rotating Layer: A Simple Case of Turbulence, {\it SCIENCE}, {\bf 208},
173-175
\item[]
Busse F H, Simitev R 2004 Inertial convection in rotating  fluid
spheres. {\it J. Fluid Mech.} {\bf 498}, 23--30.
\item[]
Christensen U R, Aubert J 2006 Scaling laws for dynamos in rotating spherical shells. {\it Geophys. J. Int.}, in press
\item[]
Christensen U R, Aubert J, Cardin P, Dormy E, Gibbons S, Glatzmaier G A, Grote E, Honkura Y, Jones C, Kono M, Matsushima M, Sakuraba A, Takahashi F, Tilgner A, Wicht J, Zhang K 2001 A numerical dynamo benchmark. {\it Phys. Earth Plan. Inter.}, {\bf 128}, 25-34
\item[]
Christensen U, Olson P, Glatzmaier, G A 1999 Numerical Modeling of the 
Geodynamo: A Systematic Parameter Study. {\it Geophys. J. Int.}, {\bf 138},
393--409.
\item[]
Chandrasekhar S 1961 {\it Hydrodynamic and Hydromagnetic Stability},
Clarendon Press, Oxford
\item[]
Childress S, Soward A M 1972 Convection-driven hydrodynamic dynamo {\it Phys. Rev. Lett.} {\bf 29}, 837-839
\item[]
Connerney J E P 1993 Magnetic Fields of the Outer Planets. {\it J. Geophys. Res.} {\bf 98}, 18659-18679 
\item[]
Crary F J, Bagenal F 1998 Remanent ferromagnetism and the interior structure of Ganymede.{\it J. Geophys. Res.} {\bf 103}, 25757-25773 
\item[] 
Dolginov Sh Sh 1977 Planetary Magnetism: A Survey {\it Geomagn. Aeron.}, {\bf
  17}, 391-406 
\item[]
Cowling T G 1934 The magnetic field of sunspots. {\it Monthly Not. Roy. astr. Soc.}{\bf 34}, 39-48
\item[]
Fearn D R 1979 Thermal and magnetic instabilities in a rapidly rotating fluid sphere.
 {\it Geophys. Astrophys. Fluid Dyn.}, {\bf 14}, 103--126 
\item[]
Giampieri G, Balogh A 2002 Mercury's thermoelectric dynamo model revisited. {\it Planet. Space Sci.}, {\bf 50}, 757-762   
\item[]
Gilbert N 1600 {\it De Magnete, Gilbert Club revised English translation}, Chiswick Press, London
\item[]
Glatzmaier G A, Roberts P H 1995 A three-dimensional self-consistent computer simulation of a geomagnetic field reversal. {\it Nature} {\bf 377}, 203-209  
\item[]
Glatzmaier G A, Roberts P H 1996 An anelastic evolutionary geodynamo simulation driven by compositional and thermal convection. {\it Physica D} {\bf 97}, 81-94  
\item[]
Grote E, Busse F H 2001 Dynamics of convection and dynamos in rotating
spherical fluid shells. {\it Fluid Dyn. Res.}, {\bf 28}, 349-368 
\item[]
Grote E,  Busse F H, Tilgner A 2000  Regular and chaotic spherical
dynamos.  {\it Phys. Earth Planet. Inter.} {\bf 117}, 259-272 
\item[]
Herzenberg A 1958 Geomagnetic Dynamos. {\it Phil. Trans. Roy. Soc. London} {\bf A250}, 543-585
\item[]
Hide R 1974 Jupiter and Saturn. {\it Proc. Roy. Soc. London} {\bf A 336}, 63-84 
\item[]
Holme R, Bloxham J 1996 The magnetic fields of Uranus and Neptun: Methods and models. {\it J. Geophys. Res.} {\bf 101}, 2177-2200
\item[]
Jacobs J A 1979 Planetary Magnetic Fields.  {\it Geophys. Res. Letts.}, {\bf 6}, 213-214
\item[]
Jones C A 2000 Convection-driven geodynamo models. {\it Phil. Trans. Roy. Soc. London} {\bf A358}, 873-897
\item[]
Jones C A 2003 Dynamos in planets. In: Thompson M J, Christensen-Daalsgard J. (eds.) {\it Stellar Astrophysical Fluid Dynamics} Cambridge University Press, pp. 159-176
\item[]
Jones C A, Soward A M,  Mussa A I 2000 The onset of thermal
convection in a rapidly rotating sphere.  {\it J. Fluid Mech.} {\bf
405}, 157-179
\item[]
Kivelson M G, Khurana K K, Russell C T, Walker R J, Warnecke J, Coroniti F V, Polanskey C, Southwood D J, Schubert G 1996 Discovery of Ganymede's magnetic field by the Galileo spacecraft. {\it Nature} {\bf 384}, 537-541
\item[]
Kutzner C,  Christensen, U R 2000 Effects of driving
mechanisms in geodynamo models. {\it Geophys. Res. Lett.} {\bf
27} 29--32. 
\item[]
Kutzner K,  Christensen U 2002 From stable dipolar towards
reversing numerical dynamos.  {\it Phys. Earth Planet. Inter.}, {\bf
  131}, 29-45 
\item[]
Larmor J 1919 How could a rotating body such as the sun become a magnet? {\it Brit. Ass. Advan. Sci. Rep.} 159-160
\item[]
Love J J 2000 Dynamo action and the nearly axisymmetric magnetic field of Saturn.  {\it Geophys. Res. Letts.}, {\bf 27}, 2889-2892
\item[]
Malkus W V R 1994 Energy Sources for Planetary Dynamos. In: Proctor M R E, Gilbert A D (eds.)  {\it Lectures on Planetary and Solar Dynamos} Cambridge University Press, pp. 161-179
\item[]
Malkus W V R 1968 Precession of the Earth as the Cause of Geomagnetism. {\it SCIENCE} {\bf 160} 259-264
\item[]
Margot J, Peale S, Jurgens R F, Slade M A, Holin I V 2004 Earth-based measurements of Mercury's forced librations in longitude. {\it Eos Trans. AGU} {\bf 85} GP33-03
\item[]
Nellis W J, Hamilton D C, Holmes N C, Radousky H B, Ree F H, Mitchell A C, and Nicol M 1988 The Nature of the Interior of Uranus Based on Studies of Planetary Ices at High Dynamic Pressure. {\it SCIENCE}, {\bf 240}, 779-781
\item[]
Nellis W J, Weir S T, and Mitchell A C 1996 Metallization and Electrical Conductivity of Hydrogen in Jupiter. {\it SCIENCE}, {\bf 273}, 936-938
\item[]
Runcorn S 1975 An ancient lunar magnetic field. {\it Nature} {\bf 253} 701-703
\item[]
Ruzmaikin A A, Starchenko S V 1991 On the origin of Uranus and Neptune magnetic fields. {\it Icarus}, {\bf 93}, 82-87
\item[]
Sarson G R, Jones C A, Zhang K, Schubert G 1997 Magnetoconvection dynamos and the magnetic field of Io and Ganymede. {\it SCIENCE}, {\bf 276}, 1106-1108
\item[]
Showman A P, Stevenson D J, Malhotra R 1997 Coupled orbital and thermal evolution of Ganymede. {\it Icarus} {\bf 129}, 367-383
\item[]
Simitev R, Busse F H 2005 Prandtl-number dependence of convection-driven 
    dynamos in rotating spherical fluid shells. {\it J. Fluid Mech.} {\bf 532}, 365-388
\item[] 
Somerville R C J, Lipps F B 1973 A Numerical Study in Three Space Dimensions of B\'enard Convection in a rotating Fluid. {\it J. Atmos. Sci.} {\bf 30}, 590-596   
\item[]
Soward A M 1974 A convection driven dynamo, I, The weak field case. {\it Phil. Trans. Roy. Soc. London} {\bf A275}, 611-645
\item[]
Stanley S, Bloxham J, Hutchison W E, and Zuber M T 2005 Thin shell dynamo models consistent with Mercury's weak observed magnetic field. {\it Earth Plan. Sci. Lett.} {\bf 234} 27-38
\item[]
Stegman D R, Jellinek A M, Zatman S A, Baumgardner J R and Richards M A  2003 An early lunar core dynamo driven by thermochemical mantle convection. {\it Nature} {\bf 421} 143-146
\item[]
Stevenson D J 1979 Turbulent Thermal Convection in the Presence of Rotation and a Magnetic Field. {\it Geophys. Astrophys. Fluid Dyn.}, {\bf 12}, 139-169
\item[]
Stevenson D J 1982 Reducing the non-axisymmetry of a planetary dynamo and an application to Saturn. {\it Geophys. Astrophys. Fluid Dyn.}, {\bf 21}, 113-127
\item[]
Stevenson D J 1983 Planetary magnetic fields. {\it Rep. Prog. Phys.} {\bf 46}, 555-620
\item[]
Stevenson D J 1984 The Energy Flux Number and Three Types of Planetary Dynamo. {\it Astron. Nachr.} {\bf 305}, 257-264
\item[]
Stevenson D J 1987 Mercury's magnetic field: a thermoelectric dynamo? {\it Earth Planet. Sci. Lett.} {\bf 82}, 114-
\item[]
Stevenson D J 2001 Mars' core and magnetism. {\it Nature} {\bf 412} 214-219
\item[]
Stevenson D J 2003 Planetary magnetic fields. {\it Earth Planet. Sci. Lett.} {\bf 208}, 1-11
\item[]
Tilgner A 1999 Spectral Methods for the Simulation of Incompressible Flows in Spherical Shells.  {\it Int. J. Numer. Meth. Fluids}, {\bf 30}, 713-724 
\item[]
Tilgner A 2005 Precession driven dynamos. {\it Phys. Fluids}, {\bf 17}, 034104-1-6 
\item[]
Tilgner A, Busse F H 1997 Finite amplitude convection in rotating 
spherical fluid shells, {\it J. Fluid Mech.}, {\bf 332}, 359-376
\item[]
Vanyo J P 1984 Earth core motions: experiments with spheroids. {\it Geophys. J. Roy. astr. Soc.} {\bf 77}, 173-183
\item[]
Zhang K 1994 On coupling between the Poincar\'e equation and the heat
equation. {\it J. Fluid Mech.}, {\bf 268}, 211-229 
\item[]
Zhang K 1995 On coupling between the Poincar\'e equation and the heat
equation: no-slip boundary condition. {\it J. Fluid Mech.}, {\bf 284},
239-256 
\item[]
\end{description}

\end{document}